\def\sorb{b}
\if s\sorb
  \documentstyle{article}
  \newlength{\absize}
  \renewcommand{\baselinestretch}{1.5}
  \renewcommand{\arraystretch}{1.5}
  
\begin{document}
  \date{}
  \pagestyle{empty}
  \thispagestyle{empty}
  \renewcommand{\thefootnote}{\fnsymbol{footnote}}
  \newcommand{\starttext}{\newpage\normalsize
    \pagestyle{plain}
    \setlength{\baselineskip}{4ex}\par
    \twocolumn\setcounter{footnote}{0}
    \renewcommand{\thefootnote}{\arabic{footnote}}}
\else
  \documentstyle[12pt]{article}
  \newlength{\absize}
  \setlength{\absize}{\textwidth}
  \renewcommand{\baselinestretch}{2.0}
  \renewcommand{\arraystretch}{2.0}
  \begin{document}
  \thispagestyle{empty}
  \pagestyle{empty}
  \renewcommand{\thefootnote}{\fnsymbol{footnote}}
  \newcommand{\starttext}{\newpage\normalsize
    \pagestyle{plain}
    \setlength{\baselineskip}{4ex}\par
    \setcounter{footnote}{0}
    \renewcommand{\thefootnote}{\arabic{footnote}}}
\fi
\newcommand{\preprint}[1]{%
  \begin{flushright}
    \setlength{\baselineskip}{3ex} #1
  \end{flushright}}
\renewcommand{\title}[1]{%
  \begin{center}
    \LARGE #1
  \end{center}\par}
\renewcommand{\author}[1]{%
  \vspace{2ex}
  {\Large
   \begin{center}
     \setlength{\baselineskip}{3ex} #1 \par
   \end{center}}}
\renewcommand{\thanks}[1]{\footnote{#1}}
\renewcommand{\abstract}[1]{%
  \vspace{2ex}
  \normalsize
  \begin{center}
    \centerline{\bf Abstract}\par
    \vspace{2ex}
    \parbox{\absize}{#1\setlength{\baselineskip}{2.5ex}\par}
  \end{center}}

\setlength{\parindent}{3em}
\setlength{\footnotesep}{.6\baselineskip}
\newcommand{\myfoot}[1]{%
  \footnote{\setlength{\baselineskip}{.75\baselineskip}#1}}
\renewcommand{\thepage}{\arabic{page}}
\setcounter{bottomnumber}{2}
\setcounter{topnumber}{3}
\setcounter{totalnumber}{4}
\newcommand{\figsize}{}
\renewcommand{\bottomfraction}{1}
\renewcommand{\topfraction}{1}
\renewcommand{\textfraction}{0}

%
\font\ddbar=ddbar
\unitlength=1mm
\def\slash#1{#1 \hskip -0.5em /}
%
%
%

\preprint{\#HUTP-92/A053 \\ 12/92 \\ hep-ph/9301212}

\vfill
\title{$D$-$\bar{D}$ Mixing in Heavy Quark
       Effective Field Theory: The Sequel}
\vfill
\author{Thorsten Ohl \\ Giulia Ricciardi \\ Elizabeth H.~Simmons \\\hfil\\
        Lyman Laboratory of Physics\thanks{electronic mail addresses:
                {\tt \{ohl,ricciardi,simmons\}@physics.harvard.edu}}\\
        Harvard University \\ Cambridge, MA 02138 }
\date{}

\vfill
\abstract{We perform a quantitative analysis of $D^0$-$\bar{D^0}$
  mixing in Heavy Quark Effective Field Theory (HqEFT) including
  leading order QCD corrections.  We find an enhancement of the
  short-distance contribution by a factor of two or three.}

\vfill

\starttext

\section{Introduction}
\label{sec:intro}

The mixing of neutral particles particle with their antiparticles through
flavor changing neutral currents is a sensitive probe  of flavor physics.
Ever since the~$K_L^0$-$K_S^0$ mass difference was used to predict the
charm quark mass~\cite{GL74},  these systems have been used to test the
standard model and to search for signs of new physics.  Recently, there has
been  interest in~$D^0$-$\bar D^0$ mixing as a window into dynamical
symmetry breaking mechanisms~\cite{Geo92a,CH92,Geo92b}.

The traditional analysis of~$D^0$-$\bar D^0$ mixing in the standard model
is plagued by large uncertainties.  While the ``short distance''
contributions are known to be small, it has been argued~\cite{Wol85,DGHT86}
that the ``dispersive'' contributions from second-order weak interactions
with mesonic intermediate states may be  considerably larger. This differs
substantially from the situations in the~$K^0$-$\bar K^0$ and~$B^0$-$\bar
B^0$ systems. In the former, the short distance contributions are expected
to be of the same size as the dispersive contribution~\cite{DGHT84}; in the
latter, dispersive effects are expected to be negligible~\cite{DGHT86}.

It has recently been argued~\cite{Geo92b} that Heavy Quark Effective Field
Theory ~(HqEFT)~\cite{HqEFT,Geo91} and naive dimensional
analysis~\cite{NDA} suggest that the dispersive contributions
are smaller than previously estimated, implying that cancellations occur
between contributions from different classes of intermediate mesonic
states.
In this paper, we perform a quantitative analysis of~$D^0$-$\bar D^0$
mixing in the HqEFT framework in order to test that assertion. We first
calculate the matching contributions at the charm quark scale where the
charm quark is removed from the effective theory, leaving only a heavy
color charge. Then, we calculate the one-loop anomalous dimensions that
contribute to the running of the operators between the charm mass
and~$\Lambda_{QCD}$.  Although the change in scale is not large, a great
many operators contribute and there is the possibility that some of them
can get a large anomalous dimension.

After describing the effective field theory framework in
section~\ref{sec:eft}, we elaborate on the matching conditions and
introduce the operator basis for the Wilson expansion  in
section~\ref{sec:matching}.  Then we
calculate the one-loop anomalous dimensions for these operators and
use the renormalization group equation to pick up the
additional contributions from the running between the charm scale
and the hadronic scale $\sim \Lambda_{QCD}$ in section~\ref{sec:running}
and~\ref{sec:anomalous-dimensions}.  Our numerical
results appear in section~\ref{sec:results} and our conclusions are
presented in section~\ref{sec:conclusions}.

\section{The Effective Field Theory}
\label{sec:eft}

We shall need to move from the full standard model at high energies to
a low-energy effective theory of neutral $D$ meson mixing.  Since the~$D^0$
meson is a~$\bar c u$ state and the CKM~(Cabibbo-Kobayashi-Maskawa)
matrix element~$V_{ub}$ is very small, we shall ignore the effects of
the third quark generation altogether. As we
pass below the scale~$\mu = M_W$,
we integrate out the heavy weak bosons
and match onto a theory with four-fermion
weak interaction operators; this and the subsequent renormalization
group running is done in the standard way and
introduces nothing surprising.   The first really interesting physics
comes in at the charm quark threshold: at this point we remove the
charm quark from our theory and replace it by a heavy color charge
in the HqEFT.  We recognize that assuming that~$m_c$ lies far enough
above~$\Lambda_{QCD}$ for HqEFT to properly
capture the charm quark's low energy behavior is questionable.  However,
we feel that the benefits of approaching the calculation in this way make
the risks worthwhile.  And, as one usually says in applying HqEFT to the
charm quark, the approximation is well-defined and the
corrections are systematically calculable.

At this point, let us remind the reader of some of the essential physics of
HqEFT.  The central idea is that a system composed of a single heavy quark
$Q$ (where ``heavy'' means $M_Q\gg\Lambda_{QCD}$) and one or more light
quarks can be described as a heavy color charge surrounded by
generic ``brown muck'' with the characteristic energy scale of the
confining QCD interactions.  The mass and velocity of the whole system are
essentially identical to those of the heavy quark; the ``brown muck''
carries only residual momenta of the order of~$\Lambda_{QCD}\ll M_Q$.
Further, if only QCD interactions are included, the velocity of the heavy
quark is constant; additional (e.g.) weak interactions must be included to
change the heavy quark velocity.

The most common application of the HqEFT draws upon the fact that the
spin and flavor of the heavy quark are decoupled from the brown muck.  This
enables one to relate decays of various~$B$ and~$D$ mesons to one
another using the HqEFT formalism.

Our reason for using the HqEFT is rather different - what interests us is
that no large momentum (e.g.~no momentum as large as~$m_c$) can be
transferred to the brown muck.  This has the surprising consequence that
below $\mu = m_c$ charm-changing non-leptonic decays of the neutral~$D$
meson are forbidden: such decays would inevitably transfer the large charm
quark momentum to light colored degrees of freedom.  In other words, no
new operators contributing to~$D^0$-$\bar D^0$ mixing appear at scales
below the charm scale!  Other than operators arising from
short-distance physics (such as $W$ exchange in the familiar box
diagrams) the only operators contributing to~$D^0$-$\bar D^0$ mixing
arise in the matching of the ordinary charm quark onto the heavy charm
quark at~$\mu = m_c$.  So if the HqEFT can really be applied to the~$c$
quark, we can calculate the size of~$D^0$-$\bar D^0$  mixing in the
standard model by computing only short distance, matching and running
effects.

We should stress some peculiar features of our effective theory. Removing
the charm quark from the theory leaves a hard momentum~$p\approx m_c >
\Lambda_{QCD}$ flowing through the Feynman diagrams for the~$D^0$-$\bar
D^0$ mixing operators. This large momentum forces the internal light quark
and gluon propagators far off shell, making it possible to match them with
local terms at the charm quark scale. However, in contrast to the~$W$, the
light quarks and gluons have not been integrated out: light quarks and
gluons with momenta of the order of~$\Lambda_{QCD}$ are still present in
our effective theory. It is also crucial to note in the context
of~$D^0$-$\bar D^0$ mixing, that there are no truly ``long-distance''
contributions to the matching at the scale~$\mu\approx m_c$.  Because we
are matching at a relatively large scale~$\mu > \Lambda_{\chi SB}$, quark
loops give a good approximation to the sum over all intermediate states.
The diagrams of the form shown
in~figs.~\ref{fig:4quark-matching},~\ref{fig:6quark-matching},
and~\ref{fig:8quark-matching} give rise to local four-quark, six-quark, and
eight-quark operators below the matching scale.

\section{Matching}
\label{sec:matching}

At the $W$ mass scale, the familiar~$\Delta C = 1$
effective hamiltonian can be written in the compact form~\cite{Geo92b}:
\begin{equation}
  H_{eff} = \frac{4\,G_F}{\sqrt2}
    \left( \bar\psi_L\gamma_\mu u_L \right)
    \left(\vec\kappa\cdot\vec\tau\right)
    \left( \bar c_L\gamma^\mu \psi_L \right)
\end{equation}
where $\psi$ and $\vec \kappa$ are respectively:
\begin{equation}
   \psi = \pmatrix{s\cr d}, \qquad
   \vec \kappa = \frac{1}{2}\pmatrix{%
                         \cos^2\theta - \sin^2\theta\cr
                         -i\cr
                         2 \cos\theta \,\sin\theta }
\end{equation}
The subscript $L$ denotes application of the left-handed projection
operator~$(1-\gamma^5)/2$; the~$\vec\tau$ are the Pauli matrices; $\theta$
is the Cabibbo mixing angle.

In this section, we will find the coefficients of the
four-, six-, and eight-quark operators generated by matching at the
charm scale. The four-quark operators come from one-loop matching;
the six- and eight-quark  operators are generated at tree-level.
Although the latter are non leading in~$1/m_c$,
naive dimensional analysis~\cite{Geo92b,NDA} shows that their matrix
elements make important contributions to $D^0$-$\bar D^0$ mixing.

\subsection{Construction of the Operator Basis}
\label{sec:basis}

We start by  constructing a basis for the multi-quark operators arising
{}from the matching and the operators that mix with them through QCD
interactions. Each operator contains two (heavy) charm quarks and an even
number of light quarks.

Let us establish some conventions for writing down our
multi-quark operators.  A~$2n$-quark operator can be written as the
product of~$n$ currents; we shall always write the current containing
a charm anti-quark
last and that containing a heavy charm quark next-to-last.  So the form of
the operators engendering $\bar{D^0} \to D^0$ transitions
will be\footnote{Following the notation
  of~\protect\cite{Geo92b,Geo91}, we distinguish between the charm
  quark annihilation operator
  \begin{displaymath}
    c_v (x) = \frac{1+\slash v}{2} e^{im_cvx} c(x)
  \end{displaymath}
  and the anti-charm quark creation operator
  \begin{displaymath}
    {\underline c}_v (x) = \frac{1-\slash v}{2} e^{-im_cvx} c(x)
  \end{displaymath}
  We have to add the effects of the hermitian conjugates
  of~(\ref{basisq-1}-\ref{basisq-3}) to get the full~$\Delta C = 2$
  contribution.}
\begin{eqnarray}
   \mbox{{\rm four-quark:}} && \left( {\bar c}_v \Gamma_1 u \right)
     \left( {\bar{\underline c}}_v \Gamma_2 u \right)
\label{basisq-1}\\
   \mbox{{\rm six-quark:}} &&  \left( {\bar\psi} \Gamma_1 u \right)
     \left( {\bar c}_v \Gamma_2 \psi \right)
     \left( {\bar{\underline c}}_v \Gamma_3 u \right)
\label{basisq-2} \\
   \mbox{{\rm eight-quark:}} && \left( {\bar \psi} \Gamma_1 u \right)
             \left( {\bar \psi} \Gamma_2 u \right)
     \left( {\bar c}_v \Gamma_3 \psi \right)
     \left( {\bar{\underline c}}_v \Gamma_4 \psi \right)
\label{basisq-3}
\end{eqnarray}
where each matrix $\Gamma$ contains the color and Dirac structure of the
associated current.  Now that we have established the positions
that the different
quarks occupy in the operators, we can move to a more compact notation
which omits the quarks altogether and retains only the $\Gamma$
matrices
\begin{equation}
\label{eq:2n-quark-operator}
  \left( \prod_{i=1}^{n-2} \bar q \Gamma_i q \right)
     \left( {\bar c}_v \Gamma_{n-1} q \right)
     \left( {\bar{\underline c}}_v \Gamma_n q \right)
    \mapsto \left( \bigotimes_{i=1}^{n-2} \Gamma_i \right)
         \otimes \Gamma_{n-1} \otimes \Gamma_n.
\end{equation}
Here $q$ stands for the appropriate light ($u$, $d$, or $s$) quark.
We also recognize that the color and Dirac parts of each~$\Gamma$
factorize.  Hence our tensor product notation can be usefully
broken down into separate color and Dirac tensor products:
\begin{displaymath}
  \left( \bigotimes_{i=1}^{n} \Gamma_i \right)
      = \left( \mathop{\hphantom{_C}\bigotimes\nolimits_C}_{i=1}^{n}
          \Gamma^C_i \right) \otimes
        \left( \mathop{\hphantom{_D}\bigotimes\nolimits_D}_{i=1}^{n}
          \Gamma^D_i \right).
\end{displaymath}
{}From here on, we distinguish the tensor products in
color ($\otimes_C$) and Dirac ($\otimes_D$) space by subscripts, and
we reserve the unqualified tensor product symbol~$\otimes$ for the tensor
product of color and Dirac space.

The factorization of the~$\Gamma$ matrices implies that the
construction of the operator basis naturally divides into two parts.
Finding a basis in color space is conceptually simple because all the
quarks have the same color properties.  One just enumerates all
singlets in the tensor product of $n$ copies of $1 \oplus (N^2 - 1)$.
We shall see that this is trivial when $n$ = 2 or 3 and is not much
more complicated for $n$ = 4.

Finding a basis in Dirac space is more interesting since the charm and
light quarks have different Dirac properties. Since we are calculating
to lowest nonvanishing order in the light quark masses, the full standard
model theory tells us that the light quarks participating
in~$D^0$-$\bar D^0$ mixing are left-handed:
\begin{equation}
\label{eq:LH}
   (1+\gamma^5) q = 0\ .
\end{equation}
According to the HqEFT, the heavy charm quarks are static colors sources
that do not have a `handedness'.  What they do have are equations of
motion that must be satisfied:
\begin{eqnarray}
\label{eq:EOM-first}
   (\slash v - 1) c_v & = & 0 \\
\label{eq:EOM-last}
   (\slash v + 1) {\underline c}_v & = & 0
\end{eqnarray}
and these will affect the Dirac structure of the operator basis.
The vector space of the Dirac operators
\begin{equation}
\label{eq:ourop}
  \left( \mathop{\hphantom{_D}\bigotimes\nolimits_D}_{i=1}^{n}
      \Gamma^D_i \right)
\end{equation}
for which we want to find a basis
is, in fact, a subspace of the $n$-fold tensor product
\begin{equation}
   {\cal G}_n = \mathop{\hphantom{_D}\bigotimes\nolimits_D}_{i=1}^{n}
         {\cal G}_1
\end{equation}
of the full Dirac algebra~${\cal G}_1$. The algebra ${\cal G}_1$ is
spanned by the usual Min\-kowski space
basis $\{{\bf1},\gamma_\mu,\sigma_{\mu\nu},\gamma_\mu\gamma^5,\gamma^5\}$.
The trick is to define precisely {\em which\/} subspace it corresponds to.
Since all
Lorentz indices in~(\ref{eq:ourop}) are fully contracted, (\ref{eq:ourop})
belongs to the subspace of~${\cal G}_n$ that transforms as a scalar
under the proper Lorentz group; we call this subspace~${\cal G}^0_n$.
Now if~(\ref{eq:ourop}) spanned~${\cal G}^0_n$, in order to
find a basis we would only need to extend the usual inner
product on~${\cal G}_1$
\begin{equation}
  \left(\Gamma,\Gamma'\right) = \frac{1}{4}
      \mathop{\rm Tr}\left(\Gamma^\dagger\Gamma'\right)
\end{equation}
to an inner product on~${\cal G}^0_n$
\begin{equation}
  \label{eq:expinner}
  \left(\Gamma,\Gamma'\right) = \prod_{i=1}^n \frac{1}{4}
      \mathop{\rm Tr}\left(\Gamma_i^\dagger\Gamma_i'\right).
\end{equation}
Then a basis of~(\ref{eq:ourop}) could be constructed by finding a
complete set orthogonal with respect to the
inner product~(\ref{eq:expinner}).
However, the peculiar properties of the currents connecting heavy and
light quarks as summarized
in~(\ref{eq:LH}) and~(\ref{eq:EOM-first}-\ref{eq:EOM-last}) mean
that the basis vectors of~${\cal G}^0_n$ whose construction we
have just described are not independent.

Now we can finish the construction of the basis for the
operators~(\ref{eq:ourop}).
We can use the right and left projection operators
\begin{eqnarray}
  \pi_r & = &
      \mathop{\hphantom{_D}\bigotimes\nolimits_D}\limits_{i=1}^{n}
                \frac{1}{2}\left(1-\gamma^5\right)\\
  \pi_l & = &
      \left( \mathop{\hphantom{_D}\bigotimes\nolimits_D}\limits_{i=1}^{n-2}
             \frac{1}{2}\left(1-\gamma^5\right) \right)
      \otimes \frac{1}{2} \left(1 + \slash v\right)
      \otimes \frac{1}{2} \left(1 - \slash v\right)
\end{eqnarray}
to project onto the subspace of operators
with non vanishing matrix elements. Then the kernel of the
combined action of these projection operators
\begin{displaymath}
  \left\{ \Gamma\in{\cal G}_n^0 \vert\pi_l \Gamma \pi_r = 0 \right\}
\end{displaymath}
is the physically irrelevant portion of ${\cal G}^0_n$.
Therefore the elements of the factor space
\begin{equation}
\label{eq:factor-space}
  \overline{{\cal G}_n}
    = {\cal G}_n^0/\left\{ \Gamma\in{\cal G}_n^0 \vert
               \pi_l \Gamma \pi_r = 0 \right\}
\end{equation}
correspond to the physically inequivalent operators in which we are
interested.  $\overline{{\cal G}_n}$ can naturally be equipped with an
inner product by
\begin{equation}
\label{eq:inner-product}
  \overline{\left(\Gamma,\Gamma'\right)}
      = \prod_{i=1}^n \frac{1}{4}
      \mathop{\rm Tr}\left(\Gamma_i^\dagger\pi_l\Gamma_i'\pi_r\right).
\end{equation}
The construction of a basis in $\overline{{\cal G}_n}$ is now
equivalent to finding a maximal subspace of ${\cal G}_n^0$, such
that~$\overline{\left(\,\cdot\,,\,\cdot\,\right)}$  is non-degenerate.

\subsection{Four-Quark Operators}
\label{sec:4quark-matching}

The basis for the four-quark operators is particularly
simple.  In color space the only possible
operators are the tensor product of a pair
of color singlets or a pair of color octets
\begin{eqnarray}
   {\tau^4}_1   & = & {\bf 1} \otimes_C {\bf 1} \\
   {\tau^4}_2   & = & T_a \otimes_C T_a.
\end{eqnarray}
The Dirac space $\overline{{\cal G}_2}$ is spanned by
\begin{eqnarray}
\label{eq:4q-Dirac-first}
   {\Upsilon^4}_1 & = & \gamma_L^\mu \otimes_D \gamma_{L,\mu} \\
\label{eq:4q-Dirac-last}
   {\Upsilon^4}_2 & = & {\bf 1}_L \otimes_D {\bf 1}_L.
\end{eqnarray}

Let us show explicitly how conditions~(\ref{eq:LH})
and~(\ref{eq:EOM-first}-\ref{eq:EOM-last}) are used to
reduce~${\cal G}^0_2$
to~(\ref{eq:4q-Dirac-first},\ref{eq:4q-Dirac-last}).
First, condition~(\ref{eq:LH}) selects the left handed
sector $\{{\bf 1}_L,$ $\gamma_{L,\mu},$ $\sigma_{L,\mu\nu}\}$
of ${\cal G}_1$.
Applying conditions~(\ref{eq:EOM-first}-\ref{eq:EOM-last}) removes
operators containing~$\slash v$ or~$v_\mu\sigma_L^{\mu\nu}$ from the
basis:
\begin{eqnarray*}
    \slash v_L \otimes_D \slash v_L & = &
   -{\Upsilon^4}_2 \\
   v_\mu \sigma_L^{\mu\nu} \otimes_D \gamma_{L,\nu} & = &
i\,{\Upsilon^4}_1 +i\, {\Upsilon^4}_2 \\
  \gamma_{L,\nu}  \otimes_D v_\mu \sigma_L^{\mu\nu}  & = &
-i\,{\Upsilon^4}_1 -i\, {\Upsilon^4}_2 \\
   v_\mu \sigma_L^{\hphantom{L,}\mu\rho} \otimes_D
      v_\nu \sigma_{L,\hphantom{\nu}\rho}^{\hphantom{L,}\nu} & = &
  {\Upsilon^4}_1 + {\Upsilon^4}_2.
\end{eqnarray*}
The self duality relation
\begin{equation}
\label{eq:sigma-self-duality}
  \sigma_L^{\mu\nu}  = - \frac{i}{2}
     \varepsilon^{\mu\nu\kappa\lambda} \sigma_{L,\kappa\lambda}
\end{equation}
makes operators containing contractions of~$\sigma_L$
and~$\varepsilon$ redundant and leads to
\begin{eqnarray}
\label{eq:sigma-spin1}
    \sigma_{L,\hphantom{\mu}\alpha}^{\hphantom{L,}\mu}
                 \otimes_D \sigma_L^{\hphantom{L}\nu\alpha}
             - \sigma_{L,\hphantom{\nu}\alpha}^{\hphantom{L,}\nu}
                 \otimes_D \sigma_L^{\hphantom{L}\mu\alpha}
       & = &  - i\varepsilon^{\mu\nu\kappa\lambda} \sigma_{L,\kappa\alpha}
                \otimes_D
                \sigma_{L,\lambda}^{\hphantom{L,\lambda}\alpha} \\
\label{eq:sigma-spin2}
    \sigma_{L,\hphantom{\mu}\alpha}^{\hphantom{L,}\mu}
                 \otimes_D \sigma_L^{\hphantom{L}\nu\alpha}
             + \sigma_{L,\hphantom{\nu}\alpha}^{\hphantom{L,}\nu}
                 \otimes_D \sigma_L^{\hphantom{L}\mu\alpha}
       & = &  \frac{1}{2} g^{\mu\nu} \sigma_L^{\kappa\lambda}
                \otimes_D \sigma_{L,\kappa\lambda}.
\end{eqnarray}
Finally, contracting~(\ref{eq:sigma-spin2}) with~$v_\mu v_\nu$ and
using~(\ref{eq:EOM-first}-\ref{eq:EOM-last}), we arrive at
\begin{displaymath}
    \sigma_L^{\kappa\lambda} \otimes_D \sigma_{L,\kappa\lambda}
       = 4\cdot\left( {\Upsilon^4}_1 + {\Upsilon^4}_2 \right)
\end{displaymath}
which leaves us with~(\ref{eq:4q-Dirac-first},\ref{eq:4q-Dirac-last})
as our basis.

\begin{figure}[tb]
  \begin{center}
  \unitlength=1mm
  \begin{picture}(100,28)
    \put(-3,3){$c$}
    \put(-3,22){$u$}
    \put(0,0){{\ddbar f}}
    \put(18,2){$d,s$}
    \put(18,22){$d,s$}
    \put(40,3){$u$}
    \put(40,22){$c$}
    \put(49,12){$\Longrightarrow$}
    \put(57,3){$c_v$}
    \put(57,22){$u$}
    \put(60,0){{\ddbar g}}
    \put(100,22){${\underline{c}}_v$}
    \put(100,3){$u$}
  \end{picture}
  \end{center}
  \caption{Matching of the four-quark operators at $\mu\approx m_c$}
  \label{fig:4quark-matching}
\end{figure}

We can now calculate the matching condition at $\mu\approx
m_c$ from the one loop diagram in fig.~\ref{fig:4quark-matching}.
\begin{eqnarray}
\label{eq:fish}
  \lefteqn{ \left( -i 4 \frac{G_F}{\sqrt 2}
                         \sin\theta\cos\theta \right)^2
            \left[ {\tau^4}_1 \otimes
        \int\!\frac{d^4l}{(2\pi)^4} \right. } \nonumber\\
       &   & \qquad
          \gamma_L^\mu \left( \frac{i}{m_c\slash v + \slash l - m_s}
             -  \frac{i}{m_c\slash v + \slash l - m_d} \right)
                 \gamma_L^\nu \nonumber\\
       &   & \qquad \left.
   \otimes_D \gamma_{L,\nu} \left( \frac{i}{\slash l - m_s}
             -  \frac{i}{\slash l - m_d} \right)
                 \gamma_{L,\mu} \right]\nonumber \\
       & = & -i \frac{1}{\pi^2} \frac{G_F^2}{2}
                  \sin^2\theta \cos^2\theta
		    \frac{m_s^4}{m_c^2} \nonumber \\
       &   & \quad \left[ {\tau^4}_1 \otimes \left\{
               2 \cdot \gamma_L^\mu \otimes_D \gamma_{L,\mu}
             + \slash v \sigma_L^{\mu\nu}
                  \otimes_D \slash v \sigma_{L,\mu\nu}
                   \right\} \right] \\
       & = & i \frac{1}{\pi^2} \frac{G_F^2}{2}
                  \sin^2\theta \cos^2\theta
		    \frac{m_s^4}{m_c^2}
                \left[ {\tau^4}_1 \otimes \left( 2{\Upsilon^4}_1
                               + 4{\Upsilon^4}_2 \right) \right] \nonumber
\end{eqnarray}
to lowest order $m_s^2/m_c^2$.

\subsection{Six-Quark Operators}
\label{sec:6quark-matching}

For the six-quark operators the color basis is
\begin{eqnarray}
  {\tau^6}_1 & = & {\bf 1} \otimes_C {\bf 1} \otimes_C {\bf 1} \\
  {\tau^6}_2 & = & T_a \otimes_C T_a \otimes_C {\bf 1} \\
  {\tau^6}_3 & = & T_a \otimes_C {\bf 1} \otimes_C T_a \\
  {\tau^6}_4 & = & {\bf 1} \otimes_C T_a \otimes_C T_a \\
  {\tau^6}_5 & = & d_{abc} \cdot T_a \otimes_C T_b \otimes_C T_c \\
  {\tau^6}_6 & = & f_{abc} \cdot T_a \otimes_C T_b \otimes_C T_c
\end{eqnarray}
and the basis for the Dirac structure is
\begin{eqnarray}
\label{eq:6quark-basis-first}
  {\Upsilon^6}_{1} & = &
    \gamma_L^\mu \otimes_D {\bf 1}_L \otimes_D \gamma_{L,\mu} \\
  {\Upsilon^6}_{2} & = &
    \gamma_L^\mu \otimes_D \gamma_{L,\mu} \otimes_D {\bf 1}_L \\
  {\Upsilon^6}_{3} & = &
    \gamma_L^\mu \otimes_D \left(
       \sigma_{L,\mu\nu} \otimes_D \gamma_L^\nu
          +  \gamma_L^\nu \otimes_D \sigma_{L,\mu\nu} \right) \\
  {\Upsilon^6}_{4} & = &
    {\slash v}_L \otimes_D \gamma_L^\mu \otimes_D \gamma_{L,\mu} \\
\label{eq:6quark-basis-last}
  {\Upsilon^6}_{5} & = &
    {\slash v}_L \otimes_D {\bf 1}_L \otimes_D {\bf 1}_L.
\end{eqnarray}
In finding the Dirac basis, we first note that
the Dirac piece of the current with two light quarks must contain
a~$\gamma^\mu_L$ since all light quarks are left-handed.  Then we apply the
steps discussed in the previous section for the four-quark case.  An
additional relation
\begin{equation}
  {\slash v} \sigma_L^{\kappa\lambda}
    = i v^\kappa \gamma_L^\lambda - i v^\lambda \gamma_L^\kappa
      + v_\mu \varepsilon^{\mu\kappa\lambda\nu} \gamma_{L,\nu}
\label{6q-additional}
\end{equation}
allows us to trade contractions of~$v$,~$\varepsilon$, and~$\gamma_L$
for $\sigma_L$
and vice versa. It also allows us to remove the antisymmetric
counterpart of~${\Upsilon^6}_{3}$ from the basis.

\begin{figure}[tb]
  \begin{center}
  \unitlength=1mm
  \begin{picture}(100,58)
    \put(-3,3){$\psi$}
    \put(-3,14){$u$}
    \put(-3,22){$c$}
    \put(0,0){{\ddbar 8}}
    \put(18,15){$d,s$}
    \put(40,3){$\psi$}
    \put(40,14){$c$}
    \put(40,22){$u$}
    \put(49,12){$\Longrightarrow$}
    \put(57,3){$\psi$}
    \put(57,14){$u$}
    \put(57,22){$c_v$}
    \put(60,0){{\ddbar 9}}
    \put(100,3){$\psi$}
    \put(100,14){${\underline c}_v$}
    \put(100,22){$u$}
    \put(-3,33){$\psi$}
    \put(-3,44){$c$}
    \put(-3,52){$u$}
    \put(0,30){{\ddbar 6}}
    \put(18,45){$d,s$}
    \put(40,33){$\psi$}
    \put(40,44){$u$}
    \put(40,52){$c$}
    \put(49,42){$\Longrightarrow$}
    \put(57,33){$\psi$}
    \put(57,44){$c_v$}
    \put(57,52){$u$}
    \put(60,30){{\ddbar 7}}
    \put(100,33){$\psi$}
    \put(100,44){$u$}
    \put(100,52){${\underline c}_v$}
  \end{picture}
  \end{center}
  \caption{Six-quark matching}
  \label{fig:6quark-matching}
\end{figure}

Now let us consider the matching at the charm mass  scale.
{}From the first graph in figure~\ref{fig:6quark-matching} we obtain
\begin{eqnarray}
\label{eq:whi}
\label{eq:6quark-matching}
\lefteqn{ \left( -i 4 \frac{G_F}{\sqrt 2} \right)^2 \cos\theta \sin\theta
   \, (\vec\kappa\cdot\vec\tau) \,\, \left[ {\tau^6}_1 \otimes \left\{
           \gamma_{L,\nu} \otimes_D
     \gamma_{L,\mu} \vphantom{\frac{1}{2}}
       \right. \right. } & & \nonumber \\
& & \left. \left. \otimes_D
 \left({\gamma_L}^\nu \frac{i}{m_c \slash v-m_s} \gamma_L^\mu-
{\gamma_L}^\nu \frac{i}{m_c\slash v-m_d} \gamma_L^\mu\right)
  \right\} \right].
\end{eqnarray}
Here the matrix $(\vec\kappa\cdot\vec\tau)$
acts on the flavor doublets $\psi$ of down-type quarks appearing in the
six-quark operator (see e.g.~equation~\ref{basisq-2})
\begin{displaymath}
  (\vec\kappa\cdot\vec\tau)^{ij} (\bar\psi^i\ u )(\bar c_v\ \psi^j)
  (\bar{\underline c}_v \ u).
\end{displaymath}
where repeated indices are summed.
Since $m_d \ll m_s$, (\ref{eq:whi}) reduces to
\begin{displaymath}
-i\,8 G_F^2 \,\cos\theta \sin\theta\,
\left(\vec\kappa\cdot\vec\tau\right) \,\, \frac{m_s^2}{m_c^3}
 \left[ {\tau^6}_1 \otimes \left\{ \gamma_{L,\nu} \otimes_D
\gamma_{L,\mu} \otimes_D
 {\gamma_L}^\nu \,
 \slash{v} \, \gamma_L^\mu \right\} \right].
  \label{match1}
\end{displaymath}
Expressing the Dirac structure in terms of our operator basis yields
\begin{equation}
-i\,8 G_F^2 \,\cos\theta \sin\theta\,
\left(\vec\kappa\cdot\vec\tau\right) \,\frac{m_s^2}{m_c^3}
 \left[ {\tau^6}_1 \otimes
   \left\{ \frac{1}{2} ({\Upsilon^6}_1 + {\Upsilon^6}_2)
      -i {\Upsilon^6}_3 - {\Upsilon^6}_4 \right\} \right].
  \label{match4}
\end{equation}
The second graph in figure~\ref{fig:6quark-matching} also gives a matching
contribution of the form~(\ref{match4}).  In this case, the flavor
indices of the
matrix $(\vec\kappa\cdot\vec\tau$) are contracted as follows
\begin{displaymath}
  (\vec\kappa\cdot\vec\tau)^{ij} (\bar\psi^i u) (\bar c_v u)
   (\bar{\underline c}_v \psi^j).
\end{displaymath}

\subsection{Eight-Quark Operators}
\label{sec:8quark-matching}

In studying the color structure for the eight-quark operators, we
found it useful to
employ a generic $SU(N)$ color group.
The number of singlets in the
tensor product of four adjoint representations of $SU(N)$ is not
entirely
independent of~$N$: there are 8 singlets for $N = 3$ and 9 for~$N > 3$.
The difference arises because the rank four tensor
\begin{equation}
\label{eq:tau24}
   d_{abe} d_{cde} +  d_{ace} d_{bde}
    +  d_{ade} d_{bce}
         - \frac{1}{N}\delta_{ab} \delta_{cd}
          - \frac{1}{N}\delta_{ac} \delta_{bd}
               - \frac{1}{N}\delta_{ad} \delta_{bc}
\end{equation}
vanishes~\cite{MSW68} for $SU(3)$ and is non-vanishing
when~$N > 3$.  We circumvent
this difficulty by constructing a basis with one element proportional
to~(\ref{eq:tau24}). Then the basis for $N =3$ is the
subspace of the $N>3$ basis that is perpendicular to
the~(\ref{eq:tau24})
direction. The 24 operators for~$N > 3$
are listed in appendix~\ref{sec:8q-basis-color}.

A straightforward extension of the methods described for the
four-quark and six-quark Dirac bases enables us to find the~14 independent
eight-quark Dirac structures listed in
appendix~\ref{sec:8q-basis-Dirac}.

\begin{figure}[tb]
  \begin{center}
  \unitlength=1mm
  \begin{picture}(100,28)
    \put(-3,5){$c$}
    \put(10,0){$\psi$}
    \put(-3,18){$u$}
    \put(0,22){$\psi$}
    \put(0,0){{\ddbar a}}
    \put(41,19){$c$}
    \put(28,23){$\psi$}
    \put(38,0){$\psi$}
    \put(41,5){$u$}
    \put(49,12){$\Longrightarrow$}
    \put(57,5){$c_v$}
    \put(70,0){$\psi$}
    \put(57,18){$u$}
    \put(70,22){$\psi$}
    \put(60,0){{\ddbar A}}
    \put(101,19){${\underline{c}}_v$}
    \put(88,23){$\psi$}
    \put(88,0){$\psi$}
    \put(101,5){$u$}
  \end{picture}
  \end{center}
  \caption{An example of a leading contribution to eight-quark matching.
    This graph corresponds to~$(l{:}l{:}{:})$ in the notation of
    section~\protect\ref{sec:one-loop-integrals}.}
  \label{fig:8quark-matching}
\end{figure}

Now we are ready to consider the matching of the eight-quark operators.
We have to study graphs similar to the one shown in
figure~\ref{fig:8quark-matching} where a gluon is exchanged between
two four-fermion operators.  Note that not all graphs give a
leading (order $\sim 1/m_c^4$) contribution in the effective theory.
For instance, graphs in which the gluon couples to two heavy quarks
give operators with coefficients of order~$ 1/m_c^6$: one factor
of~$1/m_c^2$ comes from each off-shell heavy quark propagator and one more
comes from the propagator of the gluon carrying the large momentum of
order~$m_c$. Similarly, graphs in which the gluon couples to one heavy and
one light quark make contributions of order ($1/m_c^5$) since an off-shell
light quark propagator gives a factor of ($1/m_c$).  The leading graphs are
those in which the gluon couples to two light quarks.\footnote{%
  Hence the graph in figure~5 of
  reference~\cite{Geo92b} is actually subleading.}

The contribution to the matching made by the graph in
figure~\ref{fig:8quark-matching} is
\begin{eqnarray}
\label{eq:8q-matching}
\lefteqn{  8  i \frac{G_F^2 4\pi\alpha_s (m_c)}{m_c^4}
    \left(\vec\kappa\cdot\vec\tau\right)\,
    \left(\vec\kappa\cdot\vec\tau\right) } && \nonumber \\
& & \left[ {\tau^8}_2 \otimes \left(
     -{\Upsilon^8}_{1}-{\Upsilon^8}_{2}+
 {\Upsilon^8}_{3}+i  {\Upsilon^8}_{5} -i  {\Upsilon^8}_{6}
-\frac{1}{2}  {\Upsilon^8}_{7} \right. \right. \\
& & \left.\left.+\frac{1}{2}  {\Upsilon^8}_{8}
-\frac{1}{2}  {\Upsilon^8}_{9}+ \frac{1}{2}  {\Upsilon^8}_{10}
+\frac{i}{2}  {\Upsilon^8}_{11} -\frac{i}{2}  {\Upsilon^8}_{12}
\right) \right] \nonumber.
\end{eqnarray}
Compared to the coefficient of the four-quark
operators,~(\ref{eq:8q-matching}) has an extra suppression of
order~$\alpha_s/m_c^2$, but no~$m_s^4$ suppression.
The matrices $\vec\kappa\cdot\vec\tau$
act on the flavor indices of the
doublets $\psi$ of down-type quarks.
Explicitly,
\begin{equation}
(\vec\kappa\cdot\vec\tau)^{ij}\,
(\vec\kappa\cdot\vec\tau)^{kl} (\bar\psi^i u) (\bar\psi^k u)
   (\bar c_v \psi^j) (\bar{\underline{c}}_v \psi^l).
\end{equation}
The contribution from the other graphs is
presented in appendix~\ref{sec:8q-matching-appendix}.

\section{One-loop integrals}
\label{sec:one-loop-integrals}
\label{sec:running}

We now compute the one-loop QCD running from~$m_c$
to~$\Lambda_{QCD}$.  After discussing the several types of loop integrals
occurring in the calculation, we perform the necessary algebra.
The calculations for the four-quark operators are easily performed
using standard techniques.  The eight-quark operator basis is however
large enough to make use of symbolic manipulation programs.
All our results for the six- and eight-quark operators have been
obtained and verified by two independent calculations using the
symbolic manipulation tools {\tt Form}~\cite{Ver91} and
{\tt Mathematica}$^{TM}$~\cite{Wol91}.  The algorithm of the {\tt Form}
program uses Dirac and $SU(N)$ trace techniques by treating the Dirac
algebra as a factor space as discussed in section~\ref{sec:basis}.  The
{\tt Mathematica} program on the other hand implements the reduction from
${\cal G}_n$ to $\overline{{\cal G}_n}$ explicitly,
using relations like (\ref{6q-additional}).

\begin{figure}[tb]
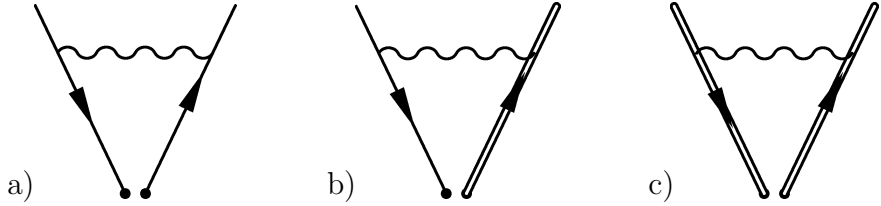

  \begin{center}
    a){\ddbar 1} \hfil b){\ddbar 2} \hfil c){\ddbar 3}
  \end{center}
  \caption{Gluon insertion into light-light, heavy-light, and
    heavy-heavy pairs of quark lines.}
  \label{fig:currents}
\end{figure}

Each loop diagram contributing to the running of our multi-quark operators
below the charm scale has one internal gluon and two internal quarks. The
form of the loop integral depends on whether the internal quarks are heavy
or light.  The three possibilities (both heavy, both light, one of each)
are sketched in figure~$\ref{fig:currents}$. Note that each quark line
internal to the loop  is clearly bounded at one by the gluon insertion and
at the other by a  vertex connecting it to another quark line.  The result
of any loop integral  may be expressed simply in terms of the color and
Dirac matrices  associated with those gluon insertions and vertices.  So
long as at  least one of the internal quarks is heavy, the loop integral
turns out  to be independent of the Dirac structure of the vertices.  For
example, (see also appendix~\ref{sec:one-loop-appendix}) the
integral corresponding to
diagram~\ref{fig:currents}b yields
\begin{equation} I^{hl} =  2 \frac{\alpha}{4\pi}
\frac{\mu^\epsilon}{\epsilon} \left[ T_a V_1\right]^C \left[ T_a
V_2\right]^C + \mbox{ finite },
\label{eq:nea}
\end{equation}
and that for diagram ~\ref{fig:currents}c
yields
\begin{equation} I^{hh} =  -4 \frac{\alpha}{4\pi}
\frac{\mu^\epsilon}{\epsilon}
\left[ T_a V_1\right]^C \left[ T_a V_2 \right]^C + \mbox{ finite }.
\label{eq:neb}
\end{equation}
Here $T_a$ is the color
matrix arising from a gluon insertion and the~$V_i$ is the color matrix
{}from the vertex at the other end of the internal quark line.   When both
internal quarks are light, however, the Dirac structure of the internal
quarks' vertices is relevant (fig.~\ref{fig:currents}a)
\begin{equation} I^{ll}  =  \frac{1}{2}
\frac{\alpha}{4\pi} \frac{\mu^\epsilon}{\epsilon} \left[ T_a V_1 \right]^C
\left[ T_a V_2 \right]^C \left[ \gamma^\mu \gamma^\nu V_1\right]^D \left[
\gamma_\mu \gamma_\nu V_2 \right]^D + \mbox{ finite }.
\label{eq:nec}
\end{equation}
What we have just seen is that the evaluation
of each loop diagram falls neatly
into three pieces: the momentum integration,  the product of color
matrices, and the product of Dirac matrices.  The first of these is
essentially done;  we shall address the matrix algebra in the
section~\ref{sec:anomalous-dimensions}.

We  can represent the sum of all 1-loop diagrams renormalizing a $2n$-quark
operator ${\cal O}_n$ as a sum over tensor products
of linear operators~$C_i$ and~$D_i$ acting on
the color~$SU(N)$ and Dirac bases
\begin{equation}
  \delta{\cal O}_n = \frac{\alpha}{4\pi} \frac{\mu^\epsilon}{\epsilon}
          \sum\limits_{i=1}^{n_D} b_i \left(
         C_i \otimes D_i \right) {\cal O}_n
\label{oneloop-dc}
\end{equation}
where~$n_D$ is the number of diagrams. The $C_i$ and $D_i$ come directly
{}from equations~\ref{eq:nea},~\ref{eq:neb} and~\ref{eq:nec}. The
possible values for the coefficients $b_i$ are $\{\pm1/2, \pm2, \pm4\}$
as we have seen.

Since we will need to evaluate quite a number of one-loop diagrams, it will
be useful to have a way of identifying individual diagrams without drawing
a picture of each one.  As all the one-loop diagrams involved in the
running of a particular multi-quark operator differ only in the placement
of the single gluon line, this identification is not difficult.  For
example, recall that we represent a generic four-quark operator
by the tensor
product~(cf.~equations~(\ref{basisq-1}-\ref{basisq-3},
\ref{eq:2n-quark-operator}))
\begin{displaymath}
\Gamma_1 \otimes \Gamma_2.
\end{displaymath}
The one-loop diagram in which a gluon is attached to the charm quark
and charm anti-quark of this operator could be represented by
marking the placement of the gluon insertions
\begin{displaymath}
G\,\Gamma_1 \otimes G\,\Gamma_2.
\end{displaymath}
The same information about the placement of the gluon insertions can be
conveyed by the briefer notation
\begin{displaymath}
  (l{:}l)
\end{displaymath}
which indicates that the gluon is attached to both the left side of
the first
current and the left side of the second current.  The one-loop diagram in
which the gluon is attached to the up quarks would be~$(r{:}r)$ in
this
notation.  The four diagrams in which the gluon is attached to one heavy
and one light quark would be represented as~$(lr{:})$, $({:}lr)$,
$(l{:}r)$, and $(r{:}l)$.  The
direct correspondence between this `colon' notation and the one-loop
diagrams for the four-quark operators is shown in
figure~\ref{fig:4quark-running}.  The extension
of the notation to the six- and eight-quark operators is
straightforward.

Now we can restate our results in colon notation:
\begin{itemize}
  \item{} From (\ref{basisq-1}-\ref{basisq-3}) we see that
each graph where a gluon connects two heavy lines is of the
form $(\ldots{:}l{:}l)$.  For such graphs, $D_i = 1$ and $b_i = 4$.
  \item{} The graphs with a light and a heavy quark connected by a gluon
 also have $D_i = 1$.  The coefficient $b_i$ always has magnitude 2.  The
sign of $b_i$ is positive if the graph is of the
form~$(\ldots r{:}\ldots{:}l{:})$ or~$(\ldots r{:}\ldots{:}l)$
and is negative if the graph has the
form~$(\ldots l{:}\ldots{:}l{:})$ or~$(\ldots l{:}\ldots{:}l)$.
  \item{} The graphs where a  gluon connects two light quarks
     can have a  non-trivial Dirac structure.  If the graph is of
     the form~$(\ldots l{:}\ldots{:}l\ldots)$
or~$(\ldots r{:}\ldots{:}r\ldots)$ then $D_i \neq 1$ and
$b_i =-{1\over2}$.  Otherwise, we can apply the following relation
for left-handed Dirac matrices $\Gamma_L$
\footnote{%
{}From
  \begin{eqnarray*}
  \lefteqn{\gamma^\mu\gamma^\nu \gamma_L^\alpha
      \otimes_D \Gamma_L \gamma_\nu\gamma_\mu -
          4 \cdot \gamma_L^\alpha \otimes_D \Gamma_L} \\
    & = &  \sigma^{\mu\nu} \gamma_L^\alpha
              \otimes_D \Gamma_L \sigma_{\mu\nu}
         = \sigma_R^{\mu\nu} \gamma_L^\alpha
              \otimes_D \Gamma_L \sigma_{L,\mu\nu} \\
    & = & \left( {\bf 1} \otimes_D \Gamma_L \right)
          \left( \sigma_R^{\mu\nu} \otimes_D \sigma_{L,\mu\nu} \right)
          \left( \gamma_L^\alpha \otimes_D {\bf 1} \right)
  \end{eqnarray*}
  we see that~(\ref{eq:trivial-lr}) is a consequence of the
  identity
  \begin{displaymath}
     \sigma_R^{\mu\nu} \otimes_D \sigma_{L,\mu\nu} = 0.
  \end{displaymath}}
\begin{equation}
\label{eq:trivial-lr}
  \gamma^\mu\gamma^\nu \gamma_L^\alpha
    \otimes_D \Gamma_L \gamma_\nu\gamma_\mu =
        4 \cdot \gamma_L^\alpha \otimes_D \Gamma_L
\label{8quarks:semplylights}
\end{equation}
to find that $D_i = 1$ and $b_i = 2$.
\end{itemize}

These observations have  interesting consequences in the
large~$N$ limit.  The only diagrams which can give a
leading contribution of order $\alpha N$ are those which
that have a trivial Dirac structure from~(\ref{eq:trivial-lr}).
This can be seen most easily in~'t~Hooft's double line
notation~\cite{tHo74}, where it is obvious that all diagrams
connecting two quark lines on the same side of the operator are not
planar and therefore subleading.  Therefore the different Dirac
structures do not mix in the limit of large~$N$.

\section{Anomalous Dimensions}
\label{sec:anomalous-dimensions}

Having calculated the divergent pieces of the loop integrals, we can
now proceed to use the
renormalization group equation
\begin{equation}
\label{eq:RGE-Gamma}
  \left( \mu\frac{\partial}{\partial\mu}
       + \beta(g) \frac{\partial}{\partial g}
       + n_l \gamma_l + n_h \gamma_h
         - \gamma_{n} \right)
       {\cal O}_{n} = 0
\end{equation}
to extract the anomalous dimensions $\gamma_{n}$.
Here $n_l$ and $n_h$ are, respectively, the number of heavy and
light quark fields in the operator~${\cal O}_n$ and $n= n_l+n_h$.
Using~\cite{Geo91}
\begin{eqnarray}
  \beta(g) & = & - \frac{g_s^3}{16\pi^2} \frac{11N - 2n_f}{3} \\
  \gamma_l & = & - C_F \frac{\alpha_s}{4\pi} \\
  \gamma_h & = & 2 C_F \frac{\alpha_s}{4\pi}
\end{eqnarray}
with~$n_f=3$ the number of light quark flavors below the charm scale,
we arrive at the anomalous dimension matrix
\begin{equation}
  \gamma_n = \frac{\alpha_s}{4\pi}
     \left[ \sum\limits_{i} b_i \,(C_i \otimes D_i)
            + (2n_h - n_l) C_F \cdot ({\bf 1} \otimes {\bf 1})
          \right]+O(\alpha_s^2)
\label{anomalous-genexpr}
\end{equation}
where $b_i$, $C_i$ and $D_i$ have been defined in
section~\ref{sec:one-loop-integrals}.

The Wilson coefficients $\eta_i(\mu)$ in the effective Hamiltonian
\begin{equation}
\label{eq:Heff}
  H_{eff} = G_F^2\, \sum_{i} \eta_i(\mu) O_i(\mu)
\end{equation}
satisfy the renormalization group equation
\begin{equation}
\label{eq:RGE-Coeff}
  \mu \frac{d}{d\mu} \eta_i(\mu)
     = - {\left(\gamma_n\right)_{ij}}^T \eta_j(\mu),
\end{equation}
where~$\gamma_n^T$ is the transpose of the anomalous dimension matrix
{}from~(\ref{eq:RGE-Gamma}). If we decompose~$\gamma_n^T$ as
\begin{equation}
  \gamma_n^T =  L_n \cdot \gamma_n^D \cdot R_n, \qquad
     L = R^{-1}
\label{eq:LDR-decomp}
\end{equation}
with~$\gamma_n^D$ a diagonal matrix and expand ~$\gamma_n^D$ in powers of
$\alpha_s$,
\begin{equation}
\gamma_n^D = \frac{\alpha_s}{4\pi} \tilde\gamma_n^D +
O(\alpha_s^2).
\end{equation}
the solution of~(\ref{eq:RGE-Coeff}) may be expressed in the form
\begin{equation}
  \eta(\mu) = L_n \cdot
     \left(\frac{\alpha_s(\mu)}{\alpha_s(m_c)}\right)^{{\displaystyle
       \frac{3\tilde\gamma_n^D}{2(11N-2n_f)}}} \cdot R_n \cdot \eta(m_c).
\label{eq:RGE-solution}
\end{equation}
In QCD with three light quark
flavors below~$m_c$, the exponent of~(\ref{eq:RGE-solution})
simplifies to~$\tilde\gamma_n^D/18$.  Since~$\alpha_s$ increases as we run
down from the charm scale, a positive exponent~$\tilde\gamma^D_n$
will result in enhancement of the coefficient~$\eta$.

In the rest of this section, we will calculate the anomalous dimensions of
the multi-quark operators responsible for~$D^0$-$\bar D^0$ mixing. We
will use
these anomalous dimensions to run our Wilson coefficients to a hadronic
scale of order $\Lambda_{QCD}$ in the next section.

\subsection{Four-Quark Operators}
\label{sec:4quark-running}

\begin{figure}[tb]
  \begin{center}
  \unitlength=1mm
  \begin{picture}(100,88)
    \put(0,12){$(l{:}l)$}
    \put(5,0){{\ddbar v}}
    \put(55,12){$(r{:}r)$}
    \put(60,0){{\ddbar w}}
    \put(0,42){$(r{:}l)$}
    \put(5,30){{\ddbar u}}
    \put(55,42){$(l{:}r)$}
    \put(60,30){{\ddbar t}}
    \put(0,72){$({:}lr)$}
    \put(5,60){{\ddbar r}}
    \put(55,72){$(lr{:})$}
    \put(60,60){{\ddbar s}}
  \end{picture}
  \end{center}
  \caption{Running of the four-quark operators}
  \label{fig:4quark-running}
\end{figure}
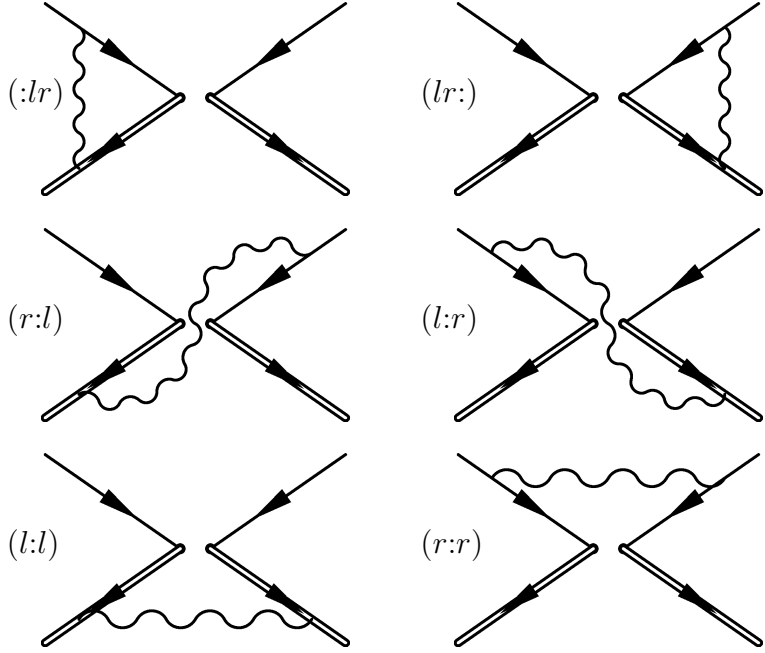

The Feynman diagrams contributing to the running of the four-quark
operators below $m_c$ are shown in fig.~\ref{fig:4quark-running}.
Using the rules and the notation explained in
section~\ref{sec:one-loop-integrals}
we can immediately infer
\begin{equation}
  D_{({:}lr)} = D_{(lr{:})} = D_{(r{:}l)}
    = D_{(l{:}r)} = D_{(l{:}l)} = {\bf 1}
\label{anomal-D}
\end{equation}
and
\begin{equation}
  b_{({:}lr)} = b_{(lr{:})} = b_{(r{:}l)} = b_{(l{:}r)} = 2,
  \quad b_{(l{:}l)} = 4, \quad b_{(r{:}r)} = - \frac{1}{2}.
\end{equation}
A short calculation yields for the color structure
\begin{eqnarray}
  C_{({:}lr)} = C_{(lr{:})} & = &
     \pmatrix{ C_F & 0 \cr 0   & - \frac{1}{2N}} \\
  C_{(r{:}l)} = C_{(l{:}r)} & = &
     \pmatrix{   0 & 1 \cr \frac{C_F}{2N} & C_F - \frac{1}{2N}} \\
  C_{(l{:}l)} = C_{(r{:}r)} & = &
     \pmatrix{   0 & 1 \cr \frac{C_F}{2N} & - \frac{1}{N}}
\end{eqnarray}
and for the remaining Dirac structure
\begin{equation}
  D_{(r{:}r)} =  \pmatrix{ 16 & 0 \cr -4 & 0}.
\label{anomaldim-Dmatrix}
\end{equation}

Using the general relation~(\ref{anomalous-genexpr}), the anomalous
dimensions for the four-quark operators read explicitly
\begin{equation}
  \left(\gamma_{4}\right)^T
    = \frac{\alpha_s}{4\pi}
     \left\{ \pmatrix{ 6C_F & \frac{4C_F}{N} \cr 8 & 6C_F - \frac{8}{N}}
              \otimes {\bf 1}
           + \pmatrix{ 0    & -\frac{C_F}{4N} \cr
                   -\frac{1}{2} & \frac{1}{2N}}
              \otimes D^T_{(r:r)}
     \right\}.
\label{eq:4q-anomal}
\end{equation}
The first term in the
sum~(\ref{eq:4q-anomal}) takes into account
the contributions of graphs ${({:}lr)}$, ${(lr{:})}$, ${(r{:}l)}$,
${(l{:}r)}$, ${(l{:}l)}$,
and the self energy subtraction.
The second  term in~(\ref{eq:4q-anomal}) is  the contribution
of the last graph~${(r{:}r)}$.

The simple form of~(\ref{eq:4q-anomal}) allows us to perform the
decomposition~(\ref{eq:LDR-decomp}) separately for the color and the
Dirac part:
\begin{eqnarray}
\label{eq:4q-anomal-diag}
  \tilde\gamma_4 & = &
      6C_F \cdot {\bf 1} \otimes {\bf 1}
     + 4 \cdot \pmatrix{ 1 - \frac{1}{N} & 0 \cr
             0 & - 1 - \frac{1}{N} }
        \otimes \pmatrix{ 0 & 0 \cr 0 & 1} \\
  L_{4} & = & \pmatrix{ 1 + \frac{1}{N} & 1 - \frac{1}{N} \cr
                                 2 & - 2 } \otimes
                       \pmatrix{ 1 & 1 \cr 0 & 4 } \\
  R_{4} & = & \pmatrix{ \frac{1}{2} & \frac{N-1}{4N} \cr
                                 \frac{1}{2} & - \frac{N+1}{4N} }
                               \otimes
                       \pmatrix{ 1 & -\frac{1}{4} \cr 0 & \frac{1}{4}}
\end{eqnarray}
and we can read off the eigenvalues from~(\ref{eq:4q-anomal-diag})
\begin{equation}
  \frac{\alpha}{\pi} \left(\frac{3}{2}C_F - 1 - \frac{1}{N}\right), \quad
  \frac{\alpha}{\pi} \left(\frac{3}{2}C_F + 1 - \frac{1}{N}\right), \quad
  \frac{\alpha}{\pi} \frac{3}{2}C_F, \quad
  \frac{\alpha}{\pi} \frac{3}{2}C_F,
\end{equation}
corresponding to the exponents $16/27$, $4/27$, $4/9$, and $4/9$
in \ref{eq:RGE-solution}.

\subsection{Six-Quark Operators}
\label{sec:6quark-running}

\begin{figure}[tb]
  \begin{center}
    {
\setlength{\unitlength}{0.240900pt}
\ifx\plotpoint\undefined\newsavebox{\plotpoint}\fi
\sbox{\plotpoint}{\rule[-0.175pt]{0.350pt}{0.350pt}}%
\begin{picture}(1300,900)(150,0)
\tenrm
\sbox{\plotpoint}{\rule[-0.175pt]{0.350pt}{0.350pt}}%
\put(242,158){\makebox(0,0)[r]{0}}
\put(244,158){\rule[-0.175pt]{4.818pt}{0.350pt}}
\put(242,284){\makebox(0,0)[r]{2}}
\put(244,284){\rule[-0.175pt]{4.818pt}{0.350pt}}
\put(242,410){\makebox(0,0)[r]{4}}
\put(244,410){\rule[-0.175pt]{4.818pt}{0.350pt}}
\put(242,535){\makebox(0,0)[r]{6}}
\put(244,535){\rule[-0.175pt]{4.818pt}{0.350pt}}
\put(242,661){\makebox(0,0)[r]{8}}
\put(244,661){\rule[-0.175pt]{4.818pt}{0.350pt}}
\put(242,787){\makebox(0,0)[r]{10}}
\put(244,787){\rule[-0.175pt]{4.818pt}{0.350pt}}
\put(264,113){\makebox(0,0){-1}}
\put(264,138){\rule[-0.175pt]{0.350pt}{4.818pt}}
\put(498,113){\makebox(0,0){-0.5}}
\put(498,138){\rule[-0.175pt]{0.350pt}{4.818pt}}
\put(733,113){\makebox(0,0){0}}
\put(733,138){\rule[-0.175pt]{0.350pt}{4.818pt}}
\put(967,113){\makebox(0,0){0.5}}
\put(967,138){\rule[-0.175pt]{0.350pt}{4.818pt}}
\put(1202,113){\makebox(0,0){1}}
\put(1202,138){\rule[-0.175pt]{0.350pt}{4.818pt}}
\put(1436,113){\makebox(0,0){1.5}}
\put(1436,138){\rule[-0.175pt]{0.350pt}{4.818pt}}
\put(264,158){\rule[-0.175pt]{282.335pt}{0.350pt}}
\put(1436,158){\rule[-0.175pt]{0.350pt}{151.526pt}}
\put(264,787){\rule[-0.175pt]{282.335pt}{0.350pt}}
\put(264,158){\rule[-0.175pt]{0.350pt}{151.526pt}}
\put(1176,158){\rule[-0.175pt]{0.350pt}{30.353pt}}
\put(1045,158){\rule[-0.175pt]{0.350pt}{15.177pt}}
\put(1019,158){\rule[-0.175pt]{0.350pt}{90.819pt}}
\put(967,158){\rule[-0.175pt]{0.350pt}{30.353pt}}
\put(941,158){\rule[-0.175pt]{0.350pt}{121.173pt}}
\put(863,158){\rule[-0.175pt]{0.350pt}{90.819pt}}
\put(811,158){\rule[-0.175pt]{0.350pt}{30.353pt}}
\put(733,158){\rule[-0.175pt]{0.350pt}{15.177pt}}
\put(707,158){\rule[-0.175pt]{0.350pt}{30.353pt}}
\end{picture}
    }
    \caption{Multiplicity distribution of the eigenvalues of the
      anomalous dimension matrix~$\tilde\gamma^D_8/18$ for the
      six-quark operators in the case~$N=3$.}
    \label{fig:multiplicity6}
  \end{center}
\end{figure}
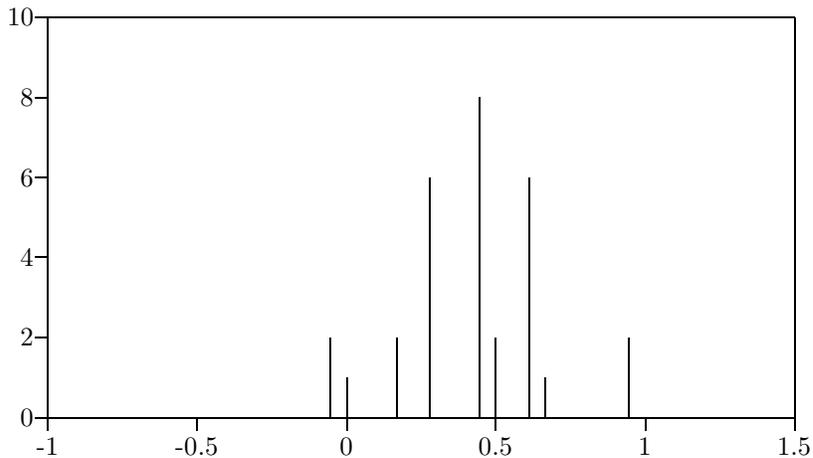

The simple structure of our results for the four-quark operators does
not persist in the six-quark case.
The gluon insertions with non trivial action on the Dirac
basis~(\ref{eq:6quark-basis-first}-\ref{eq:6quark-basis-last})
are
\begin{eqnarray}
  D_{({:}r{:}r)} & = &
    \pmatrix{
      2&2&-2i&4&0\cr
      2&2&-2i&-4&0\cr
      12i&12i&12&0&0\cr
      0&0&0&16&0\cr
      0&0&0&-4&0}
    \label{6quarks::prima}\\
  D_{(r{:}r{:})} & = &
    \pmatrix{
      6&-2&2i&-4&0\cr
      0&16&0&0&0\cr
      -18i&6i&6&12i&0\cr
      -6&-6&-2i&4&0\cr
      0&4&0&0&0}
    \label{6quarks::seconda}\\
  D_{(r{:}{:}r)} & = &
    \pmatrix{
      16&0&0&0&0\cr
      -2&6&2i&4&0\cr
      6i&-18i&6&-12i&0\cr
      6&6&2i&4&0\cr
      -4&0&0&0&0}.
    \label{6quarks::terza}
\end{eqnarray}
While each matrix has the same set of eigenvalues~$(16,16,0,0,0)$,
the corresponding eigenspaces differ and the diagonalization
method of the preceding
section is not applicable.  We can nevertheless pick a particular
value for~$N$ ($N=3$ being the natural choice) and perform the
decomposition~($\ref{eq:LDR-decomp}$) of~$\gamma_6^T$ numerically.  In
this way we obtain the multiplicities for the exponents shown in
figure~\ref{fig:multiplicity6}. The two largest exponents
are~$17/18$, resulting in an appreciable enhancement of the related
operators.

\subsection{Eight-Quark Operators}
\label{sec:8quark-running}

In the case of the eight-quark operators, there are
seven non-trivial 14-dimensional matrices describing the 1-loop
mixing of the Dirac space operators.  Rather than quoting those here, we
proceed immediately to an account of the numerical results.
In figure~\ref{fig:multiplicity} we have plotted the multiplicities of
the exponents of the anomalous dimension
matrix~$\tilde\gamma^D_8/18$ for the eight-quark operators for $N=3$.

\begin{figure}[tb]
  \begin{center}
    {
\setlength{\unitlength}{0.240900pt}
\ifx\plotpoint\undefined\newsavebox{\plotpoint}\fi
\sbox{\plotpoint}{\rule[-0.175pt]{0.350pt}{0.350pt}}%
\begin{picture}(1300,900)(150,0)
\tenrm
\sbox{\plotpoint}{\rule[-0.175pt]{0.350pt}{0.350pt}}%
\put(242,158){\makebox(0,0)[r]{0}}
\put(244,158){\rule[-0.175pt]{4.818pt}{0.350pt}}
\put(242,248){\makebox(0,0)[r]{5}}
\put(244,248){\rule[-0.175pt]{4.818pt}{0.350pt}}
\put(242,338){\makebox(0,0)[r]{10}}
\put(244,338){\rule[-0.175pt]{4.818pt}{0.350pt}}
\put(242,428){\makebox(0,0)[r]{15}}
\put(244,428){\rule[-0.175pt]{4.818pt}{0.350pt}}
\put(242,517){\makebox(0,0)[r]{20}}
\put(244,517){\rule[-0.175pt]{4.818pt}{0.350pt}}
\put(242,607){\makebox(0,0)[r]{25}}
\put(244,607){\rule[-0.175pt]{4.818pt}{0.350pt}}
\put(242,697){\makebox(0,0)[r]{30}}
\put(244,697){\rule[-0.175pt]{4.818pt}{0.350pt}}
\put(242,787){\makebox(0,0)[r]{35}}
\put(244,787){\rule[-0.175pt]{4.818pt}{0.350pt}}
\put(264,113){\makebox(0,0){-1}}
\put(264,138){\rule[-0.175pt]{0.350pt}{4.818pt}}
\put(498,113){\makebox(0,0){-0.5}}
\put(498,138){\rule[-0.175pt]{0.350pt}{4.818pt}}
\put(733,113){\makebox(0,0){0}}
\put(733,138){\rule[-0.175pt]{0.350pt}{4.818pt}}
\put(967,113){\makebox(0,0){0.5}}
\put(967,138){\rule[-0.175pt]{0.350pt}{4.818pt}}
\put(1202,113){\makebox(0,0){1}}
\put(1202,138){\rule[-0.175pt]{0.350pt}{4.818pt}}
\put(1436,113){\makebox(0,0){1.5}}
\put(1436,138){\rule[-0.175pt]{0.350pt}{4.818pt}}
\put(264,158){\rule[-0.175pt]{282.335pt}{0.350pt}}
\put(1436,158){\rule[-0.175pt]{0.350pt}{151.526pt}}
\put(264,787){\rule[-0.175pt]{282.335pt}{0.350pt}}
\put(264,158){\rule[-0.175pt]{0.350pt}{151.526pt}}
\put(1375,158){\rule[-0.175pt]{0.350pt}{8.672pt}}
\put(1306,158){\rule[-0.175pt]{0.350pt}{21.681pt}}
\put(1288,158){\rule[-0.175pt]{0.350pt}{8.672pt}}
\put(1271,158){\rule[-0.175pt]{0.350pt}{13.009pt}}
\put(1236,158){\rule[-0.175pt]{0.350pt}{8.672pt}}
\put(1219,158){\rule[-0.175pt]{0.350pt}{4.336pt}}
\put(1202,158){\rule[-0.175pt]{0.350pt}{13.009pt}}
\put(1167,158){\rule[-0.175pt]{0.350pt}{39.026pt}}
\put(1150,158){\rule[-0.175pt]{0.350pt}{13.009pt}}
\put(1132,158){\rule[-0.175pt]{0.350pt}{60.707pt}}
\put(1115,158){\rule[-0.175pt]{0.350pt}{30.353pt}}
\put(1097,158){\rule[-0.175pt]{0.350pt}{13.009pt}}
\put(1080,158){\rule[-0.175pt]{0.350pt}{26.017pt}}
\put(1063,158){\rule[-0.175pt]{0.350pt}{86.483pt}}
\put(1045,158){\rule[-0.175pt]{0.350pt}{13.009pt}}
\put(1028,158){\rule[-0.175pt]{0.350pt}{69.379pt}}
\put(1011,158){\rule[-0.175pt]{0.350pt}{47.698pt}}
\put(993,158){\rule[-0.175pt]{0.350pt}{43.362pt}}
\put(976,158){\rule[-0.175pt]{0.350pt}{69.379pt}}
\put(959,158){\rule[-0.175pt]{0.350pt}{73.715pt}}
\put(941,158){\rule[-0.175pt]{0.350pt}{13.009pt}}
\put(924,158){\rule[-0.175pt]{0.350pt}{138.517pt}}
\put(906,158){\rule[-0.175pt]{0.350pt}{56.371pt}}
\put(889,158){\rule[-0.175pt]{0.350pt}{13.009pt}}
\put(872,158){\rule[-0.175pt]{0.350pt}{43.362pt}}
\put(854,158){\rule[-0.175pt]{0.350pt}{112.500pt}}
\put(837,158){\rule[-0.175pt]{0.350pt}{21.681pt}}
\put(820,158){\rule[-0.175pt]{0.350pt}{69.379pt}}
\put(802,158){\rule[-0.175pt]{0.350pt}{26.017pt}}
\put(785,158){\rule[-0.175pt]{0.350pt}{13.009pt}}
\put(768,158){\rule[-0.175pt]{0.350pt}{26.017pt}}
\put(750,158){\rule[-0.175pt]{0.350pt}{60.707pt}}
\put(715,158){\rule[-0.175pt]{0.350pt}{60.707pt}}
\put(698,158){\rule[-0.175pt]{0.350pt}{17.345pt}}
\put(681,158){\rule[-0.175pt]{0.350pt}{21.681pt}}
\put(646,158){\rule[-0.175pt]{0.350pt}{13.009pt}}
\put(611,158){\rule[-0.175pt]{0.350pt}{8.672pt}}
\put(559,158){\rule[-0.175pt]{0.350pt}{8.672pt}}
\put(542,158){\rule[-0.175pt]{0.350pt}{8.672pt}}
\end{picture}
    }
    \caption{Multiplicity distribution of the eigenvalues of the
      anomalous dimension matrix~$\tilde\gamma^D_8/18$ for the
      eight-quark operators in the case~$N=3$.}
    \label{fig:multiplicity}
  \end{center}
\end{figure}

The largest exponents are~$37/27$ and~$11/9$ with multiplicities~$2$
and~$5$, respectively. Over all, eighteen operator coefficients have
exponents
greater than or equal to 1; for these operator coefficients, the
QCD running effectively removes the suppression factor
of~$\alpha_s(m_c)$
introduced in the charm-scale matching.

\section{Matrix Elements}
\label{sec:results}

We can now run our Wilson coefficients all the way down to a typical
hadronic scale where it is possible to evaluate the matrix
elements contributing to the $D$-meson mass difference:
\begin{equation}
  \Delta m = 2 \cdot
    \left\langle D^0 \left| H_{eff} \right\vert \bar D^0 \right\rangle.
\end{equation}
In principle, the evaluation of these matrix elements requires
non-per\-tur\-ba\-tive methods, e.g.~lattice gauge theory
calculations.  Instead,
we will estimate the size of the matrix elements using Naive Dimensional
Analysis~(NDA)~\cite{NDA,Geo92b}.

Throughout this paper, we have abbreviated our multi-quark operators by
writing down tensor-products of color and Dirac matrices without the
accompanying quark fields.  Now that we are about to calculate matrix
elements, we remind the reader that the operators~$\tau^{2n}_i$
and~$\Upsilon^{2n}_i$
do really carry quark fields with them.  For example, where we write
\begin{displaymath}
(\vec\kappa \cdot \vec\tau)\, {\tau^6}_2 \otimes {\Upsilon^6}_1
\end{displaymath}
we mean
\begin{displaymath}
  \left(\kappa \cdot \tau\right)^{mn}
  \left(\bar\psi^m \left[ T_a\otimes\gamma^\mu_L \right] u\right)\,
  \left(\bar c_v \left[ \vphantom{\gamma^\mu_L}
       T_a\otimes {\bf 1}  \right] \psi^n\right)\,
  \left(\bar c_v \left[  \vphantom{\gamma^\mu_L}
      {\bf 1}\otimes \gamma_{L,\mu}  \right] u\right).
\end{displaymath}
We will continue to use this convention in order to keep the expressions
for the matrix elements relatively compact.

The contribution of the four-quark operators to the
$D$-meson mass difference is given by the results of our
matching~(\ref{eq:fish}) and running (\ref{eq:RGE-solution})
calculations as
\begin{eqnarray}
  \left(\Delta m\right)_4
   & = & 4 \times 2 \cdot
      \frac{1}{16\pi^2} G_F^2 \sin^2\theta \cos^2\theta\,
             \frac{m_s^4}{m_c^2} \times \\
   &   &    \sum_{i,j=1,2} \eta^4_{ij}(\Lambda)
      \left\langle D^0 \left| 4 {\tau^4}_i {\Upsilon^4}_j
                \right\vert \bar D^0 \right\rangle, \nonumber
\end{eqnarray}
where~$\Lambda$ is a typical hadronic scale which we fix by the
condition~$\alpha_s(\Lambda)=1$.  Note that the factor of 4 in
the~$4{\Upsilon^4}_j$ confers a conventional normalization on the Dirac
operators (i.e.~muliplies each left-handed projection operator
by a factor of two).
The initial conditions at the charm scale may be read from~(\ref{eq:fish})
\begin{eqnarray}
  \eta^4_{11}(m_c) & = & 1 \\
  \eta^4_{12}(m_c) & = & 2 \\
  \eta^4_{21}(m_c) & = & \eta^4_{22}(m_c) = 0
\end{eqnarray}
Using the values~\cite{PDG}
\begin{eqnarray}
  \cos\theta & = & 0.975 \\
  G_F & = & 1.166 \cdot 10^{-5} \mathop{\rm GeV}\nolimits^{-2} \\
  m_c & = & 1.5 \mathop{\rm GeV},
\end{eqnarray}
we find
\begin{eqnarray}
\label{eq:delta-m-propto-ME}
  \left(\Delta m\right)_4
     & = & 23\times 10^{-17} \mathop{\rm GeV}\nolimits^{-2}\,
      \left(\frac{m_s}{0.2 \mathop{\rm GeV}}\right)^4\times \\
  & &   \sum_{i,j=1,2} \eta^4_{ij}(\Lambda)
      \left\langle D^0 \left| 4 {\tau^4}_i {\Upsilon^4}_j
                \right\vert \bar D^0 \right\rangle. \nonumber
\end{eqnarray}

We must now estimate the magnitude of the hadronic matrix elements.  NDA
tells us
\begin{equation}
  \left\langle D^0 \left| 4 {\tau^4}_i {\Upsilon^4}_j
        \right\vert \bar D^0 \right\rangle
    \approx m_c f_D^2 \approx \Lambda_{\chi SB} f^2
         \approx \frac{1}{16\pi^2} \Lambda_{\chi SB}^3
\end{equation}
where~$\Lambda_{\chi SB} \approx 1\mathop{\rm GeV}$ is the chiral symmetry
breaking scale. Note that NDA does not tell us whether the matrix
elements of our several operators will
interfere constructively or destructively.  In the context of
dimensional analysis, it is not unreasonable to assume statistical
independence of the various contributions and to add the coefficients
in quadrature.  We can also obtain a
reasonable upper estimate by adding the magnitudes of the individual
contributions.

Using these two procedures as limiting cases, the result without the
leading order QCD corrections is
\begin{displaymath}
  \left(\Delta m\right)_4^0
    \approx (0.3-0.4) \cdot 10^{-17} \mathop{\rm GeV}\,
      \left(\frac{m_s}{0.2 \mathop{\rm GeV}}\right)^4,
\end{displaymath}
in agreement with the result for the short
distance contribution from the box diagram.  Using
\begin{equation}
  \alpha_s(m_c) = 0.4
\end{equation}
the renormalization group running enhances this to
\begin{equation}
  \left(\Delta m\right)_4
    \approx (0.5-0.9) \cdot 10^{-17} \mathop{\rm GeV}\,
      \left(\frac{m_s}{0.2 \mathop{\rm GeV}}\right)^4.
\end{equation}

To evaluate the contribution of the six-quark operators to the mass
difference, we follow a similar line of reasoning.  Since NDA
suggests that the matrix element of
the $U$-spin vector operator~$\bar\psi\vec\kappa\cdot\vec\tau\psi$ is of
the order~$\sin\theta\cos\theta\,m_s f^2$~\cite{Geo92b}, we expect the
matrix elements of the six-quark operators to yield
\begin{equation}
 \left\langle D^0 \left| 8 (\vec\kappa\cdot\vec\tau){\tau^6}_i
 {\Upsilon^6}_j \right\vert\bar D^0\right\rangle
    \approx
      (\sin\theta\cos\theta\, m_s f^2)(m_c f_D^2).
\end{equation}
Combining this estimate with our results from the
matching~(\ref{eq:6quark-matching}) and running~(\ref{eq:RGE-solution})
calculations (keeping in mind that two
distinct diagrams contribute to the matching)  yields
\begin{eqnarray}
 \left(\Delta m\right)_6
   & = & 2 \cdot \frac{1}{2} \frac{\Lambda_{\chi SB}^2}{m_s
m_c}\times 2\cdot \frac{1}{16\pi^2} G_F^2 \sin^2\theta \cos^2\theta
    \frac{m_s^4}{m_c^2} \times \\
& &       \frac{1}{\sin\theta\cos\theta\,m_s f^2}
   \sum_{i,j} \eta^6_{ij}(\Lambda)
      \left\langle D^0 \left|(\vec\kappa\cdot\vec\tau)
         8 {\tau^6}_i {\Upsilon^6}_j
                \right\vert \bar D^0 \right\rangle \nonumber.
\end{eqnarray}
Then we find
\begin{displaymath}
 \left(\Delta m\right)_6^0
    \approx (0.4-0.7) \cdot 10^{-17} \mathop{\rm GeV}
      \left(\frac{m_s}{0.2 \mathop{\rm GeV}}\right)^3
\end{displaymath}
from the charm-quark matching and
\begin{equation}
 \left(\Delta m\right)_6
    \approx (0.7-2.0) \cdot 10^{-17} \mathop{\rm GeV}
      \left(\frac{m_s}{0.2 \mathop{\rm GeV}}\right)^3
\end{equation}
including the leading order QCD corrections

Repeating this reasoning for the eight-quark operators, we have
\begin{equation}
 \left\langle D^0 \left| (\vec\kappa\cdot\vec\tau)^2
16 {\tau^8}_i {\Upsilon^8}_j \right\vert \bar D^0 \right\rangle
    \approx
    (\sin\theta\cos\theta\, m_s f^2)^2 (m_c f^2_D)
\end{equation}
and applying this to~(\ref{eq:8q-matching})
and~(\ref{eq:RGE-solution}) gives
\begin{eqnarray}
  \left(\Delta m\right)_8
   & = & \frac{1}{4} \frac{\alpha(m_c)}{4\pi}
\frac{\Lambda_{\chi SB}^4}{m_s^2 m_c^2} \times 2 \cdot
   \frac{1}{16\pi^2} G_F^2 \sin^2\theta \cos^2\theta
    \frac{m_s^4}{m_c^2} \times \\
& &   \left( \frac{1}{\sin\theta\cos\theta\,m_s f^2}\right)^2
      \sum_{i,j} \eta^8_{ij}(\Lambda)
      \left\langle D^0 \left| (\vec\kappa\cdot\vec\tau)^2
       16 {\tau^8}_i {\Upsilon^8}_j
       \right\vert \bar D^0 \right\rangle \nonumber.
\end{eqnarray}
The charm-scale matching gives
\begin{displaymath}
 \left(\Delta m\right)_8^0
    \approx (0.04-0.2) \cdot 10^{-17} \mathop{\rm GeV}
      \left(\frac{m_s}{0.2 \mathop{\rm GeV}}\right)^2
\end{displaymath}
and the numerical solution of the renormalization group equation
yields
\begin{equation}
 \left(\Delta m\right)_8
    \approx (0.07-0.6) \cdot 10^{-17} \mathop{\rm GeV}
      \left(\frac{m_s}{0.2 \mathop{\rm GeV}}\right)^2.
\end{equation}
The large number of contributing operators causes a substantial
uncertainty in the last expression because our estimate of the error
{}from the unknown phases scales roughly
like~$\sqrt{\mbox{\rm number of operators}}$.  This uncertainty does
not, however, apply to the upper bound.

Adding all contributions, our final result is
\begin{equation}
 \left(\Delta m\right)_{\mathop{HqEFT}}
    \approx (0.9-3.5) \cdot 10^{-17} \mathop{\rm GeV}
\end{equation}
for~$m_s = 0.2\mathop{\rm GeV}$.

\section{Conclusions}
\label{sec:conclusions}

We have calculated the leading order QCD corrections to~$D^0$-$\bar
D^0$ mixing in the Heavy Quark Effective Field Theory.  We find that
the renormalization group running enhances~$\Delta m$ by a factor of
two to three.  While the precision of our numerical results is
limited by our incomplete knowledge of the hadronic matrix elements,
we do {\em not\/} see any
large correction to the purely short distance contribution in
the framework of HqEFT.  We therefore conclude that the
cancellations among the dispersive channels conjectured
in~\cite{Geo92b} are not removed by the leading order QCD corrections.

\section*{Acknowledgements}

We are grateful to Howard Georgi for stimulating our interest
by sharing his unconventional views on the subject
and for comments on the manuscript.
Mike Dugan provided us with valuable group theoretical suggestions.
This research has been supported in part by
the National Science Foundation under Grant \#PHY-8714654 and by the
Texas National Research Laboratory Commission under grant RGFY9206.
T.~O.~acknowledges financial support from Deutsche
Forschungsgemeinschaft~(Germany) under Grant~\#~Oh~56/1-1.
G.~R.~acknowledges financial support from INFN~(Italy).


\appendix
\section{Eight-Quark Operator Basis}
\label{sec:8q-basis}
\subsection{Color Structure}
\label{sec:8q-basis-color}

Let us fix  the notation for the invariant
tensors~$f$ and~$d$
\begin{eqnarray}
  f_{abc} & = & - 2i \mathop{\rm Tr}
      \left( \left[ T_a, T_b \right] T_c \right) \\
  d_{abc} & = & 2 \mathop{\rm Tr}
      \left( \left\{ T_a, T_b \right\} T_c \right).
\end{eqnarray}
Then the color basis for the eight-quark operators is
\begin{eqnarray}
\label{eq:SU(n)-8q-first}
  {\tau^8}_1 & = &
     {\bf 1} \otimes_C {\bf 1} \otimes_C {\bf 1} \otimes_C {\bf 1} \\
  {\tau^8}_2 & = &
     T_a \otimes_C T_a \otimes_C {\bf 1} \otimes_C {\bf 1} \\
  {\tau^8}_3 & = &
     {\bf 1} \otimes_C {\bf 1} \otimes_C T_a \otimes_C T_a \\
  {\tau^8}_4 & = &
     T_a \otimes_C {\bf 1} \otimes_C T_a \otimes_C {\bf 1} \\
  {\tau^8}_5 & = &
     {\bf 1} \otimes_C T_a \otimes_C {\bf 1} \otimes_C T_a \\
  {\tau^8}_6 & = &
     T_a \otimes_C {\bf 1} \otimes_C {\bf 1} \otimes_C T_a \\
  {\tau^8}_7 & = &
     {\bf 1} \otimes_C T_a \otimes_C T_a \otimes_C {\bf 1} \\
  {\tau^8}_8 & = &
     f_{abc} \cdot T_a \otimes_C T_b \otimes_C T_c \otimes_C {\bf 1} \\
  {\tau^8}_9 & = &
     d_{abc} \cdot T_a \otimes_C T_b \otimes_C T_c \otimes_C {\bf 1} \\
  {\tau^8}_{10} & = &
     f_{abc} \cdot T_a \otimes_C T_b \otimes_C {\bf 1} \otimes_C T_c \\
  {\tau^8}_{11} & = &
     d_{abc} \cdot T_a \otimes_C T_b \otimes_C {\bf 1} \otimes_C T_c \\
  {\tau^8}_{12} & = &
     f_{abc} \cdot T_a \otimes_C {\bf 1} \otimes_C T_b \otimes_C T_c \\
  {\tau^8}_{13} & = &
     d_{abc} \cdot T_a \otimes_C {\bf 1} \otimes_C T_b \otimes_C T_c \\
  {\tau^8}_{14} & = &
     f_{abc} \cdot {\bf 1} \otimes_C T_a \otimes_C T_b \otimes_C T_c \\
  {\tau^8}_{15} & = &
     d_{abc} \cdot {\bf 1} \otimes_C T_a \otimes_C T_b \otimes_C T_c \\
  {\tau^8}_{16} & = &
     \delta_{ab} \delta_{cd}
      \cdot T_a \otimes_C T_b \otimes_C T_c \otimes_C T_d \\
  {\tau^8}_{17} & = &
     d_{abe} d_{cde} \cdot T_a \otimes_C T_b
                 \otimes_C T_c \otimes_C T_d \\
  {\tau^8}_{18} & = &
     f_{abe} d_{cde} \cdot T_a \otimes_C T_b
                 \otimes_C T_c \otimes_C T_d \\
  {\tau^8}_{19} & = &
     d_{abe} f_{cde} \cdot T_a \otimes_C T_b
                 \otimes_C T_c \otimes_C T_d \\
  {\tau^8}_{20} & = &
     f_{abe} f_{cde} \cdot T_a \otimes_C T_b
                 \otimes_C T_c \otimes_C T_d \\
  {\tau^8}_{21} & = &
     \left( f_{ace} d_{bde} - f_{ade} d_{bce}
            - f_{bce} d_{ade} + f_{bde} d_{ace} \right) \\
      & &  \mbox{}
      \cdot T_a \otimes_C T_b \otimes_C T_c \otimes_C T_d
          \nonumber \\
  {\tau^8}_{22} & = &
     \left( \delta_{ac} \delta_{bd} - \delta_{ad} \delta_{bc} \right)
      \cdot T_a \otimes_C T_b \otimes_C T_c \otimes_C T_d \\
  {\tau^8}_{23} & = &
     \left( \delta_{ac} \delta_{bd} +  \delta_{ad} \delta_{bc} \right)
      \cdot T_a \otimes_C T_b \otimes_C T_c \otimes_C T_d \\
  {\tau^8}_{24} & = &
     \left( \vphantom{\frac{1}{N}} d_{abe} d_{cde} + d_{ace} d_{bde}
            + d_{ade} d_{bce} \right. \\
      & & \left. \mbox{}
            - \frac{1}{N} \left(
              \delta_{ab} \delta_{cd} + \delta_{ac} \delta_{bd}
               + \delta_{ad} \delta_{bc} \right) \right)
      \cdot T_a \otimes_C T_b \otimes_C T_c \otimes_C T_d \nonumber
\label{eq:SU(n)-8q-last}
\end{eqnarray}

\subsection{Dirac structures}
\label{sec:8q-basis-Dirac}

Our basis for the Dirac structure of the eight-quark  operators is
\begin{eqnarray}
  {\Upsilon^8}_{1} & = &
     \gamma_L^\mu \otimes_D \gamma_{L,\mu}
     \otimes_D \gamma_L^\nu \otimes_D \gamma_{L,\nu} \\
  {\Upsilon^8}_{2} & = &
     \gamma_L^\mu \otimes_D \gamma_L^\nu
     \otimes_D \gamma_{L,\mu} \otimes_D \gamma_{L,\nu} \\
  {\Upsilon^8}_{3} & = &
     \gamma_L^\mu \otimes_D \gamma_L^\nu
     \otimes_D \gamma_{L,\nu} \otimes_D \gamma_{L,\mu} \\
  {\Upsilon^8}_{4} & = &
     \gamma_L^\mu \otimes_D \gamma_{L,\mu}
     \otimes_D {\bf 1}_L \otimes_D {\bf 1}_L \\
  {\Upsilon^8}_{5} & = &
     \gamma_L^\mu \otimes_D \gamma_L^\nu
     \otimes_D \sigma_{L,\mu\nu} \otimes_D {\bf 1}_L \\
  {\Upsilon^8}_{6} & = &
     \gamma_L^\mu \otimes_D \gamma_L^\nu
     \otimes_D {\bf 1}_L \otimes_D \sigma_{L,\mu\nu} \\
  {\Upsilon^8}_{7} & = &
     \gamma_L^\mu \otimes_D {\slash v}_L
      \otimes_D \gamma_{L,\mu} \otimes_D {\bf 1}_L \\
  {\Upsilon^8}_{8} & = &
      {\slash v}_L \otimes_D \gamma_L^\mu
      \otimes_D \gamma_{L,\mu} \otimes_D {\bf 1}_L \\
  {\Upsilon^8}_{9} & = &
     \gamma_L^\mu \otimes_D {\slash v}_L
      \otimes_D {\bf 1}_L \otimes_D \gamma_{L,\mu} \\
  {\Upsilon^8}_{10} & = &
      {\slash v}_L \otimes_D \gamma_L^\mu
       \otimes_D {\bf 1}_L \otimes_D \gamma_{L,\mu} \\
  {\Upsilon^8}_{11} & = &
     \gamma_L^\mu \otimes_D {\slash v}_L
      \otimes_D \left( \gamma_L^\nu \otimes_D \sigma_{L,\mu\nu}
              + \sigma_{L,\mu\nu}  \otimes_D \gamma_L^\nu \right) \\
  {\Upsilon^8}_{12} & = &
      {\slash v}_L \otimes_D \gamma_L^\mu
      \otimes_D \left( \gamma_L^\nu \otimes_D \sigma_{L,\mu\nu}
              + \sigma_{L,\mu\nu}  \otimes_D \gamma_L^\nu \right) \\
  {\Upsilon^8}_{13} & = &
         {\slash v}_L \otimes_D {\slash v}_L
           \otimes_D \gamma_L^\mu \otimes_D \gamma_{L,\mu} \\
  {\Upsilon^8}_{14} & = &
     {\slash v}_L \otimes_D {\slash v}_L
         \otimes_D {\bf 1}_L \otimes_D {\bf 1}_L
\end{eqnarray}
Note that potential contributions proportional
to~$\varepsilon^{\alpha\beta\gamma\delta}
  \gamma_{L,\alpha}\otimes_D\gamma_{L,\beta}
  \otimes_D\gamma_{L,\gamma}\otimes_D\gamma_{L,\delta}$ have been
eliminated,
because the identity\footnote{%
  In a system of coordinate with $v = (\sqrt{v^2},\vec0)$
  equation~(\ref{eq:epsilon-relation})  is nothing
  but the familiar expansion formula for a determinant in terms of its
  minors.}
\begin{equation}
\label{eq:epsilon-relation}
 \forall v^2>0:
  v^2 \varepsilon^{\alpha\beta\gamma\delta}
   = v^\alpha v_\mu \varepsilon^{\mu\beta\gamma\delta}
     + v^\beta v_\mu \varepsilon^{\mu\alpha\gamma\delta}
     + v^\gamma v_\mu \varepsilon^{\mu\alpha\beta\delta}
     + v^\delta v_\mu \varepsilon^{\mu\alpha\beta\gamma}
\end{equation}
can be used to relate them to elements of the basis.

\subsection{Matching}
\label{sec:8q-matching-appendix}

In this appendix we give the results of the leading order matching for
each graph in terms of the basis
listed in appendix~\ref{sec:8q-basis}.  To distinguish the individual
graphs, we use the notation introduced at the end
of section~\protect\ref{sec:one-loop-integrals}.
All the following  terms   have the overall factor
\begin{equation}
   8  i \frac{G_F^2 4\pi\alpha_s (m_c)}{m_c^4} (\vec\kappa\cdot\vec\tau)
   (\vec\kappa\cdot\vec\tau)
\end{equation}
where $\alpha_s$ is the strong coupling and the two matrices
$\vec\kappa\cdot\vec\tau $ are intended to act over the doublets
$\psi$, as explained in section~\ref{sec:8quark-matching}
\begin{eqnarray}
  ({:}{:}r{:}r) & \mapsto &
    {\tau^8}_3 \otimes \left(
     -{\Upsilon^8}_{1}-  {\Upsilon^8}_{2}+
 {\Upsilon^8}_{3}-i  {\Upsilon^8}_{5} +i  {\Upsilon^8}_{6}
+\frac{1}{2}  {\Upsilon^8}_{7} \right. \nonumber \\
& & \left.-\frac{1}{2}  {\Upsilon^8}_{8}
+\frac{1}{2}  {\Upsilon^8}_{9}- \frac{1}{2}  {\Upsilon^8}_{10}
-\frac{i}{2}  {\Upsilon^8}_{11} +\frac{i}{2}  {\Upsilon^8}_{12}
\right) \label{matching:8q-first} \\
  (r{:}{:}{:}r) & \mapsto &
    {\tau^8}_6 \otimes \left(
     +{\Upsilon^8}_{1}-  {\Upsilon^8}_{2}
-{\Upsilon^8}_{3}+i  {\Upsilon^8}_{5} -i  {\Upsilon^8}_{6}
-\frac{1}{2}  {\Upsilon^8}_{7} \right. \nonumber \\
& & \left.+\frac{1}{2}  {\Upsilon^8}_{8}
-\frac{1}{2}  {\Upsilon^8}_{9}+ \frac{1}{2}  {\Upsilon^8}_{10}
+\frac{i}{2}  {\Upsilon^8}_{11}-\frac{i}{2}  {\Upsilon^8}_{12}
\right) \\
  (r{:}l{:}{:}) & \mapsto &
    {\tau^8}_2 \otimes \left(
     -{\Upsilon^8}_{1}+  {\Upsilon^8}_{2}
+{\Upsilon^8}_{3} -2{\Upsilon^8}_{4}+i  {\Upsilon^8}_{5} +i
{\Upsilon^8}_{6}
-\frac{1}{2}  {\Upsilon^8}_{7} \right. \nonumber \\
& & \left.+\frac{7}{2}  {\Upsilon^8}_{8}
-\frac{1}{2}  {\Upsilon^8}_{9}- \frac{1}{2}  {\Upsilon^8}_{10}
+\frac{i}{2}  {\Upsilon^8}_{11}+\frac{i}{2}  {\Upsilon^8}_{12}
+2 {\Upsilon^8}_{13}
\right) \\
  (l{:}r{:}{:}) & \mapsto &
    {\tau^8}_2 \otimes \left(
     -{\Upsilon^8}_{1}+  {\Upsilon^8}_{2}
+{\Upsilon^8}_{3} -2{\Upsilon^8}_{4}-i  {\Upsilon^8}_{5} -i
{\Upsilon^8}_{6}
+\frac{1}{2}  {\Upsilon^8}_{7} \right. \nonumber \\
& & \left.+\frac{1}{2}  {\Upsilon^8}_{8}
-\frac{7}{2}  {\Upsilon^8}_{9}+ \frac{1}{2}  {\Upsilon^8}_{10}
-\frac{i}{2}  {\Upsilon^8}_{11}-\frac{i}{2}  {\Upsilon^8}_{12}
+2 {\Upsilon^8}_{13}
\right) \\
  ({:}l{:}r{:}) & \mapsto &
    {\tau^8}_7 \otimes \left(
     {\Upsilon^8}_{1}+  {\Upsilon^8}_{2}
-{\Upsilon^8}_{3} +2{\Upsilon^8}_{4}-i  {\Upsilon^8}_{5} -i
{\Upsilon^8}_{6}
+\frac{1}{2}  {\Upsilon^8}_{7} \right. \nonumber \\
& & \left.-\frac{7}{2}  {\Upsilon^8}_{8}
+\frac{1}{2}  {\Upsilon^8}_{9}+ \frac{1}{2}  {\Upsilon^8}_{10}
-\frac{i}{2}  {\Upsilon^8}_{11}-\frac{i}{2}  {\Upsilon^8}_{12}
-2 {\Upsilon^8}_{13}
\right) \\
  (l{:}{:}{:}r) & \mapsto &
    {\tau^8}_6 \otimes \left(
     {\Upsilon^8}_{1}+  {\Upsilon^8}_{2}
-{\Upsilon^8}_{3} +2{\Upsilon^8}_{4}+i  {\Upsilon^8}_{5} +i
{\Upsilon^8}_{6}
-\frac{1}{2}  {\Upsilon^8}_{7} \right. \nonumber \\
& & \left.-\frac{1}{2}  {\Upsilon^8}_{8}
+\frac{7}{2}  {\Upsilon^8}_{9}- \frac{1}{2}  {\Upsilon^8}_{10}
+\frac{i}{2}  {\Upsilon^8}_{11}+\frac{i}{2}  {\Upsilon^8}_{12}
-2 {\Upsilon^8}_{13}
\right) \\
  (l{:}l{:}{:}) & \mapsto &
    {\tau^8}_2 \otimes \left(
     -{\Upsilon^8}_{1}-{\Upsilon^8}_{2}+
 {\Upsilon^8}_{3}+i  {\Upsilon^8}_{5} -i  {\Upsilon^8}_{6}
-\frac{1}{2}  {\Upsilon^8}_{7} \right. \nonumber \\
& & \left.+\frac{1}{2}  {\Upsilon^8}_{8}
-\frac{1}{2}  {\Upsilon^8}_{9}+ \frac{1}{2}  {\Upsilon^8}_{10}
+\frac{i}{2}  {\Upsilon^8}_{11} -\frac{i}{2}  {\Upsilon^8}_{12}
\right)
\end{eqnarray}
Two additional graphs are connected to the preceding by symmetry;
the graph $({:}r{:}r{:})$ is the same as $(r{:}{:}{:}r)$ but
with a different color factor~${\tau^8}_7 $
and~$(r{:}r{:}{:}) $ is the same as~$({:}{:}r{:}r)$ but
with a color factor ${\tau^8}_2 $.

\section{One-Loop Integrals}
\label{sec:one-loop-appendix}
\label{sec:gluon-insertions}
\label{sec:integrals}

Using the one-loop integrals
\begin{eqnarray}
  \mu^\epsilon\,
   \int\!\frac{d^{4-\epsilon}l}{(2\pi)^{4-\epsilon}}\,
                 \frac{l^\mu}{(l^2)^2lv}
           & = & 2iv^\mu \frac{1}{16\pi^2} \frac{\mu^\epsilon}{\epsilon}
             + \mbox{ finite } \\
  \mu^\epsilon\,
   \int\!\frac{d^{4-\epsilon}l}{(2\pi)^{4-\epsilon}}\,
                 \frac{1}{l^2(lv)^2}
           & = & -4i \frac{1}{16\pi^2} \frac{\mu^\epsilon}{\epsilon}
             + \mbox{ finite } \\
  \mu^\epsilon\,
   \int\!\frac{d^{4-\epsilon}l}{(2\pi)^{4-\epsilon}}\,
                 \frac{l^\mu l^\nu}{(l^2)^3}
          & = & \frac{i}{2} g^{\mu\nu} \frac{1}{16\pi^2}
              \frac{\mu^\epsilon}{\epsilon}
             + \mbox{ finite },
\end{eqnarray}
we can reduce the evaluation of our one-loop diagrams to algebra.
The gluon insertion into a light-light
current~(cf.~\ref{fig:currents}a) corresponds to the
generic expression
\begin{eqnarray}
  I^{ll} & = & \mu^\epsilon\,
         \int\!\frac{d^{4-\epsilon}l}{(2\pi)^{4-\epsilon}}\,
         \frac{-ig_{\mu\nu}\delta_{ab}}{l^2}
               \left[ -ig T_a \right]^C_1
               \left[ \gamma^\mu \frac{i}{\slash l} \right]^D_1
               \left[ -ig T_b \right]^C_2
               \left[ \gamma^\nu \frac{i}{\slash l} \right]^D_2
                                      \nonumber\\
\label{eq:light-light-current}
         & = & -i g^2 \mu^\epsilon\, \left[ T_a \right]^C_1
                      \left[ T_a \right]^C_2
            \left[ \gamma^\mu \gamma^\alpha \right]^D_1
            \left[ \gamma_\mu \gamma^\beta \right]^D_2
               \int\!\frac{d^{4-\epsilon}l}{(2\pi)^{4-\epsilon}}\,
                       \frac{l_\alpha l_\beta}{(l^2)^3} \\
  & = & \frac{1}{2} \frac{\alpha}{4\pi} \frac{\mu^\epsilon}{\epsilon}
                      \left[ T_a \right]^C_1
                      \left[ T_a \right]^C_2
            \left[ \gamma^\mu \gamma^\nu \right]^D_1
            \left[ \gamma_\mu \gamma_\nu \right]^D_2
               + \mbox{ finite }, \nonumber
\end{eqnarray}
where the notation $\left[S\right]^D_i$ refers to the insertion
of the string $S$ of Dirac matrices from right to left, starting at
the appropriate vertex;  similarly for $\left[S\right]^C_i$ in color
space.  $I_{ll}$ has to be multiplied by $-1$ for each fermion line in
which the charge flows in the opposite direction of the momentum.

The corresponding calculation for the heavy-light
current~(cf.~\ref{fig:currents}b) yields
\begin{eqnarray}
  I^{hl} & = & \mu^\epsilon\,
              \int\!\frac{d^{4-\epsilon}l}{(2\pi)^{4-\epsilon}}\,
               \frac{-ig_{\mu\nu}\delta_{ab}}{l^2}
               \left[ -ig T_a \right]^C_h
               \left[ v^\mu \frac{i}{lv} \right]^D_h
               \left[ -ig T_b \right]^C_l
               \left[ \gamma^\nu \frac{i}{\slash l} \right]^D_l
                                      \nonumber\\
\label{eq:heavy-light-current}
       & = & -ig^2 \mu^\epsilon\,\left[ T_a \right]^C_h
               \left[ T_a \right]^C_l
               \left[ \slash v \gamma^\alpha \right]^D_l
               \int\!\frac{d^{4-\epsilon}l}{(2\pi)^{4-\epsilon}}\,
               \frac{l_\alpha}{(l^2)^2(lv)} \\
       & = & 2 \frac{\alpha}{4\pi} \frac{\mu^\epsilon}{\epsilon}
               \left[ T_a \right]^C_h
               \left[ T_a \right]^C_l
               + \mbox{ finite }, \nonumber
\end{eqnarray}
where we have used $\slash v\slash v = 1$ in the last equality.
Again, $I_{hl}$ has to be multiplied by $-1$ for each fermion line in
which the charge flows in the opposite direction of the
momentum.  There is a superficial ambiguity in the sign for heavy
anti-quark lines.  One possible rule is to treat them like
anti-quarks; an equivalent one~\cite{Geo91} is to treat
them like quarks with the opposite color charge.

Finally the heavy-heavy current
(cf.~\ref{fig:currents}c) gives the contribution
\begin{eqnarray}
  I^{hh} & = & mu^\epsilon\,
          \int\!\frac{d^{4-\epsilon}l}{(2\pi)^{4-\epsilon}}\,
             \frac{-ig_{\mu\nu}\delta_{ab}}{l^2}
               \left[ -ig T_a \right]^C_1
               \left[ v^\mu \frac{i}{lv} \right]^D_1
               \left[ -ig T_b \right]^C_2
               \left[ v^\nu \frac{i}{lv} \right]^D_2
                                      \nonumber\\
\label{eq:heavy-heavy-current}
       & = & -ig^2 \mu^\epsilon \,\left[ T_a \right]^C_1
               \left[ T_a \right]^C_2
               \int\!\frac{d^{4-\epsilon}l}{(2\pi)^{4-\epsilon}}\,
               \frac{1}{(l^2)(lv)^2} \\
       & = & -4 \frac{\alpha}{4\pi} \frac{\mu^\epsilon}{\epsilon}
               \left[ T_a \right]^C_1
               \left[ T_a \right]^C_2
               + \mbox{ finite }, \nonumber
\end{eqnarray}
with the same rules for additional signs.

\end{document}